\documentclass[preprint,tightenlines,superscriptaddress,floatfix,letterpaper,nofootinbib]{revtex4-1}

\pdfoutput=1

\usepackage[justification=raggedright, margin=5mm,font={sf,small},labelfont=bf, format=plain]{caption}

\begin{document}

\begin{titlepage}

\title{
Four revolutions in physics and the second quantum revolution\\
-- a unification of force and matter by quantum information
}

\author{Xiao-Gang Wen}
\affiliation{Department of Physics, Massachusetts Institute of
Technology, Cambridge, Massachusetts 02139, USA}

\begin{abstract}
Newton's mechanical revolution unifies the motion of planets in the sky and
falling of apple on earth.  Maxwell's electromagnetic revolution  unifies
electricity, magnetism, and light.  Einstein's relativity revolution unifies
space with time, and gravity with space-time distortion.  The quantum
revolution unifies particle with waves, and energy with frequency.  Each of
those revolution changes our world view.  In this article, we will describe a
revolution that is happening now: the second quantum revolution which unifies
matter/space with information.  In other words, the new world view suggests
that elementary particles (the bosonic force particles and fermionic matter
particles)  all originated from quantum information (qubits): they are
collective excitations of an entangled qubit ocean that corresponds to our
space.  The beautiful geometric Yang-Mills gauge theory and the strange Fermi
statistics of matter particles now have a common algebraic quantum
informational origin.

\end{abstract}


\maketitle

\end{titlepage}

{\small \setcounter{tocdepth}{1} \tableofcontents }

~

~

\noindent
\emph{Symmetry is beautiful and rich.\\
Quantum entanglement is even more beautiful and richer.}

~

\section{Four revolutions in physics}

We have a strong desire to understand everything from a single or very few
origins. Driven by such a desire, physics theories were developed through the
cycle of discoveries, unification, more discoveries, bigger unification.  Here,
we would like review the development of physics and its four
revolutions\footnote{Here we do not discuss the revolution for thermodynamical
and statistical physics.}. We will see that the history of physics can be
summarized into three stages: 1) all matter is formed by particles; 2) the
discovery of wave-like matter; 3) particle-like matter = wave-like matter.  It
appears that we are now entering into the fourth stage: matter and space =
information (qubits), where qubits emerge as the origin of everything
\cite{FNN8035,Wqoem,LWqed,GW1290,W1281,W1301,YX14124784,ZW150802595}.  In the
current standard model for the elementary particles, we introduce a gauge field
for each of the electromagnetic, weak, and strong interactions.  We introduce
an anti-commuting field for each of the fermionic matter particles.  In this
fourth stage, it appears that we do not need to introduce so many different
gauge fields and anti-commuting fields to describe the bosonic force particles
and fermionic matter particles.  All we need are qubits which form a qubit
ocean that corresponds to our space.  All elementary bosonic force particles
and fermionic matter particles can appear as the collective excitations in such
a qubit ocean.  In other words, all elementary particles can be unified by
quantum information (qubits).

\subsection{Mechanical revolution}

\begin{figure}[tb] 
\centerline{ 
\includegraphics[height=2in]{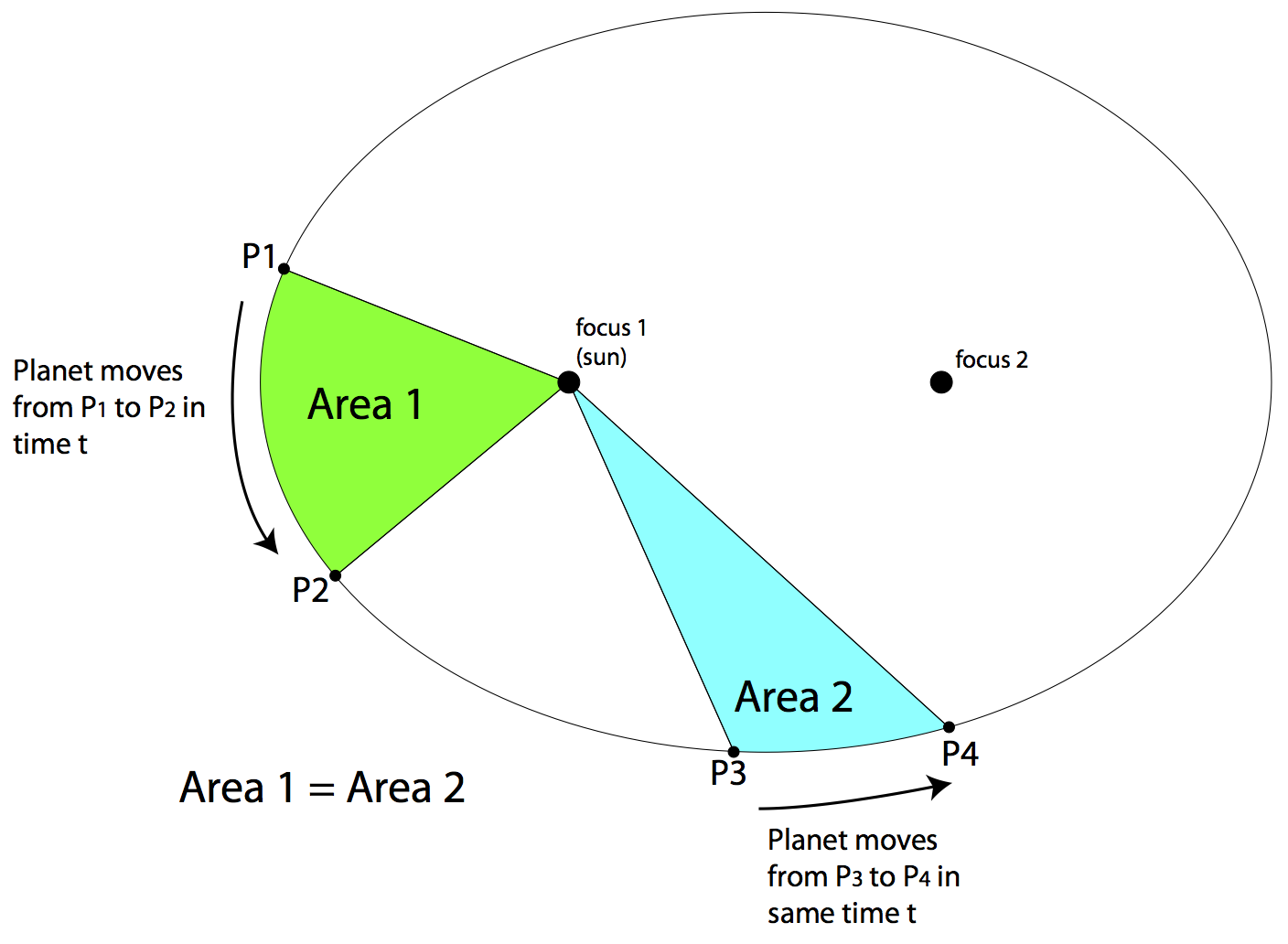} } 
\caption{
Kepler's Laws of Planetary Motion: 1) The orbit of a planet is an ellipse with
the Sun at one of the two foci.  2) A line segment joining a planet and the Sun
sweeps out equal areas during equal intervals of time. 3) The square of the
orbital period of a planet is proportional to the cube of the semi-major axis
of its orbit. 
} 
\label{Kepler2Diagram} 
\end{figure}

\begin{figure}[tb] 
\centerline{ 
\includegraphics[height=0.8in]{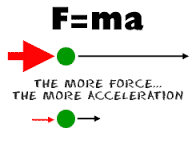}  ~~~~~~~~~
\includegraphics[height=1.in]{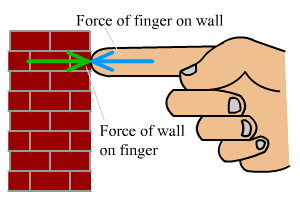}  ~~~~~~~~~
\includegraphics[height=1.in]{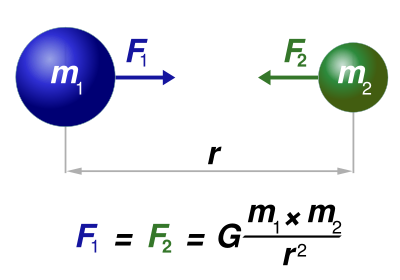}
} 
\centerline{
	(a) ~~~~~~~~~~~~~~~~~~~~~~~~~~~~~~~~~~~~~~~~~~~~~~
	(b) ~~~~~~~~~~~~~~~~~~~~~~~~~~~~~~~~~~~~~~~~~~~~~~
	(c) 
}
\caption{
	Newton laws: (a) the more force the more acceleration, no force
	no  acceleration. (b) action force = reaction force.
	(c) Newton's universal gravitation:
$F=G\frac{m_1m_2}{r^2}$, where $G=6.674\times 10^{-11} 
\frac{\text{m}^3}{\text{kg\ s}^2}$.
} 
\label{newtonlaw} 
\end{figure}

\begin{figure}[tb] 
\centerline{ \includegraphics[height=2.5in]{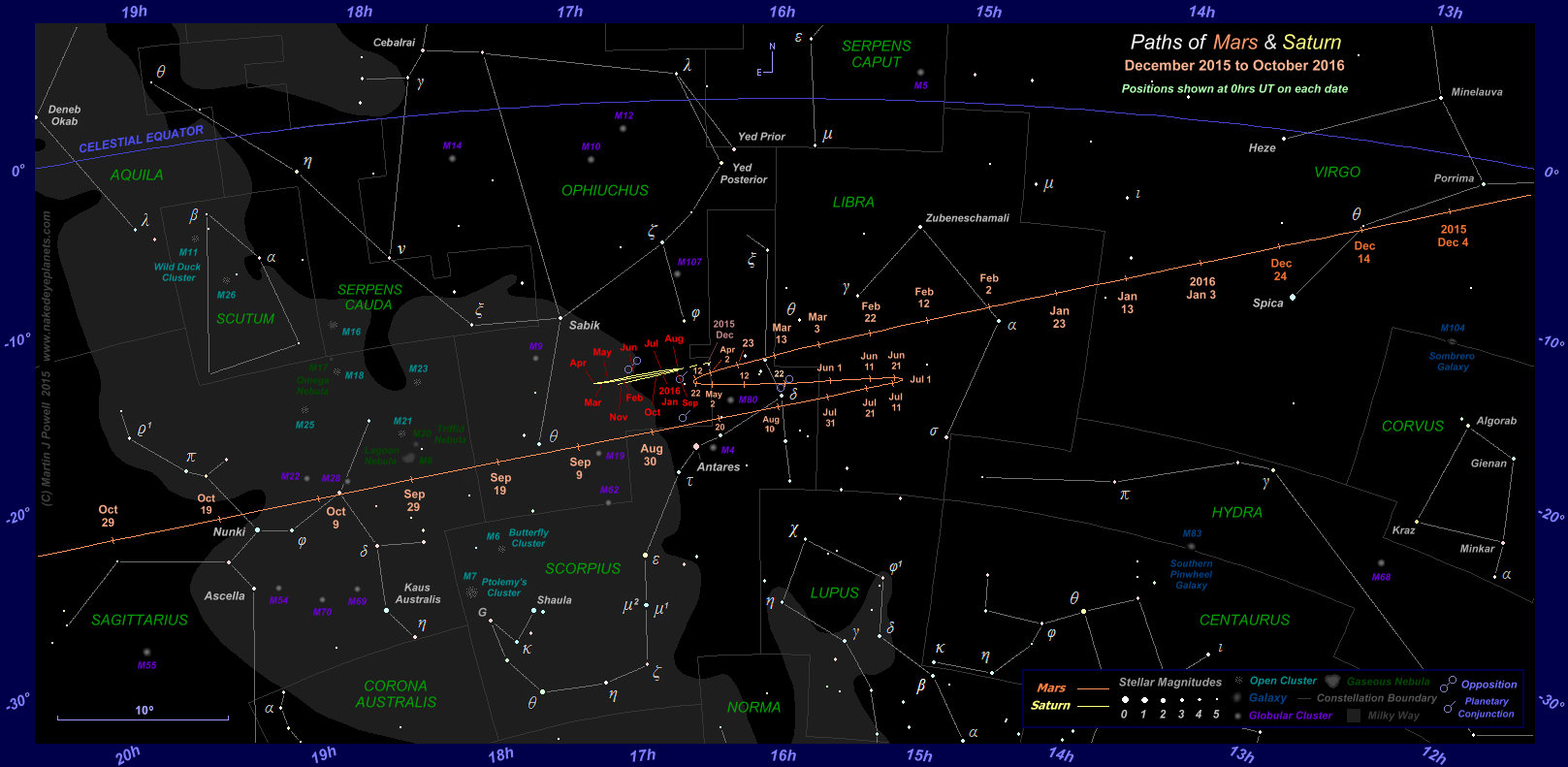}
~~
\includegraphics[height=2.0in]{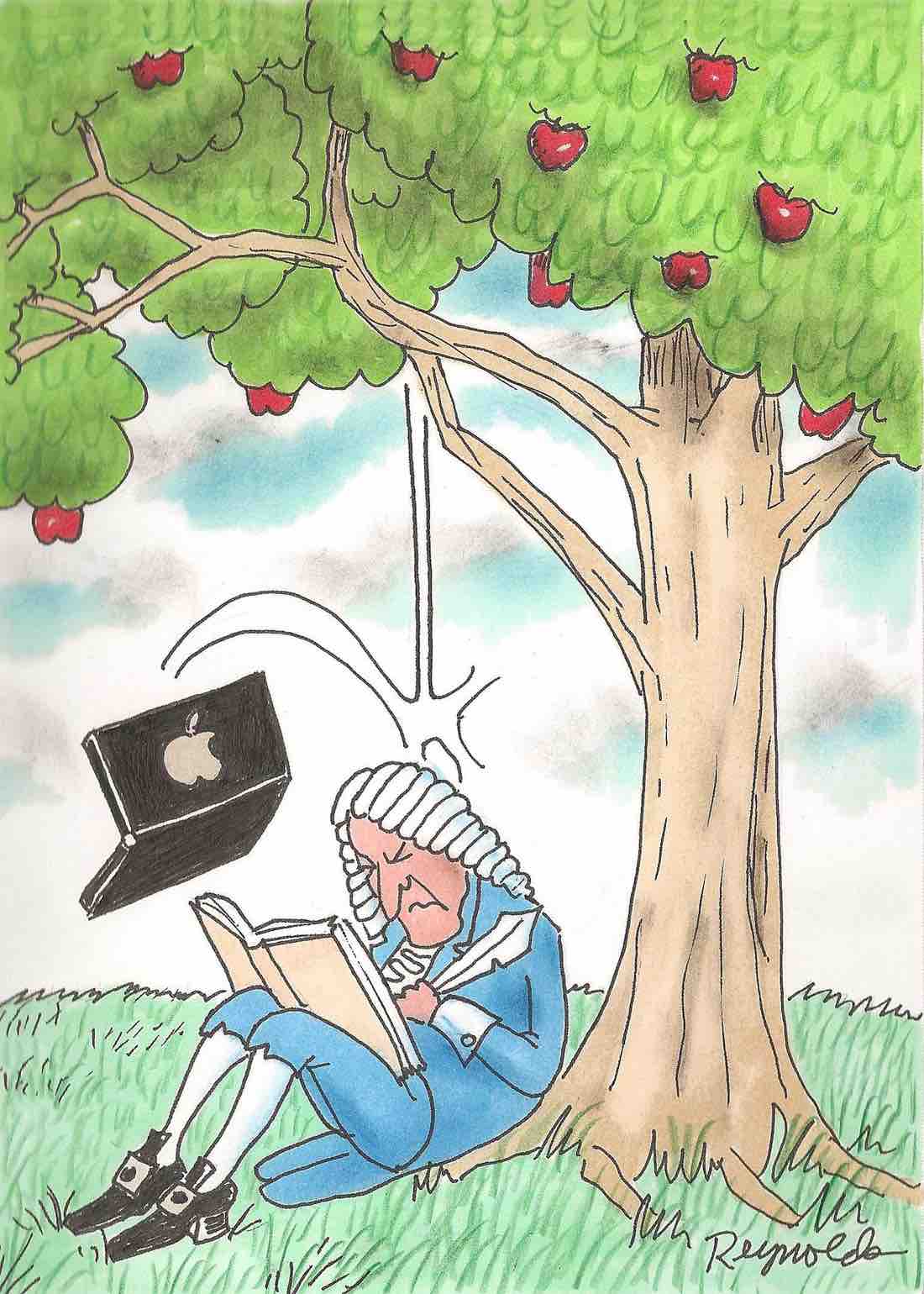}}
\caption{
The perceived trajectories of planets (Mar and Saturn) in the sky.
The falling of apple on earth and the motion of planet 
in the sky are unified by Newton theory. 
} 
\label{mars-path-dec-2015-oct-2016} 
\end{figure}

Although the down pull by the earth was realized even before human
civilization, such a phenomena did not arose any curiosity.  On the other hand
the planet motion in the sky has arose a lot of curiosity and led to many
imaginary fantasies.  However, only after Kepler found that planets move in a
certain particular way described by a mathematical formula (see Fig.
\ref{Kepler2Diagram}), people started to wonder: Why are planets so rational?
Why do they move in such a peculiar and precise way.  This motivated Newton to
develop his theory of gravity and his laws of mechanical motion (see Fig.
\ref{newtonlaw}).  
Newton's theory not only explains the planets motion, it also explains the
down-pull that we feel on earth.  The planets motion in the sky and the apple
falling on earth look very different (see Fig.
\ref{mars-path-dec-2015-oct-2016}), however, Newton's theory unifies the two
seemingly unrelated phenomena.  This is the first revolution in physics -- the
mechanical revolution.

\myfrm{
\begin{center}
\textbf{Mechanical revolution}
\\
All matter are formed by particles, which obey Newton's laws.
Interactions are instantaneous over distance.
\end{center}
}

After Newton, we view all matter as formed by particles, and use Newton's laws
for particles to understand the motion of all matter.  The success and the
completeness of Newton's theory gave us a sense that we understood everything.

\subsection{Electromagnetic revolution}

\begin{figure}[tb] 
\centerline{ 
	\includegraphics[height=1.in]{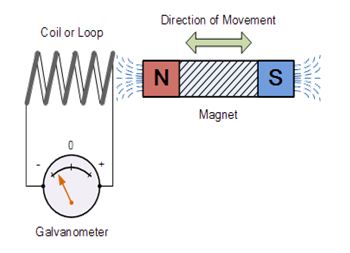}  ~~~~~~~~~~
	\includegraphics[height=1.in]{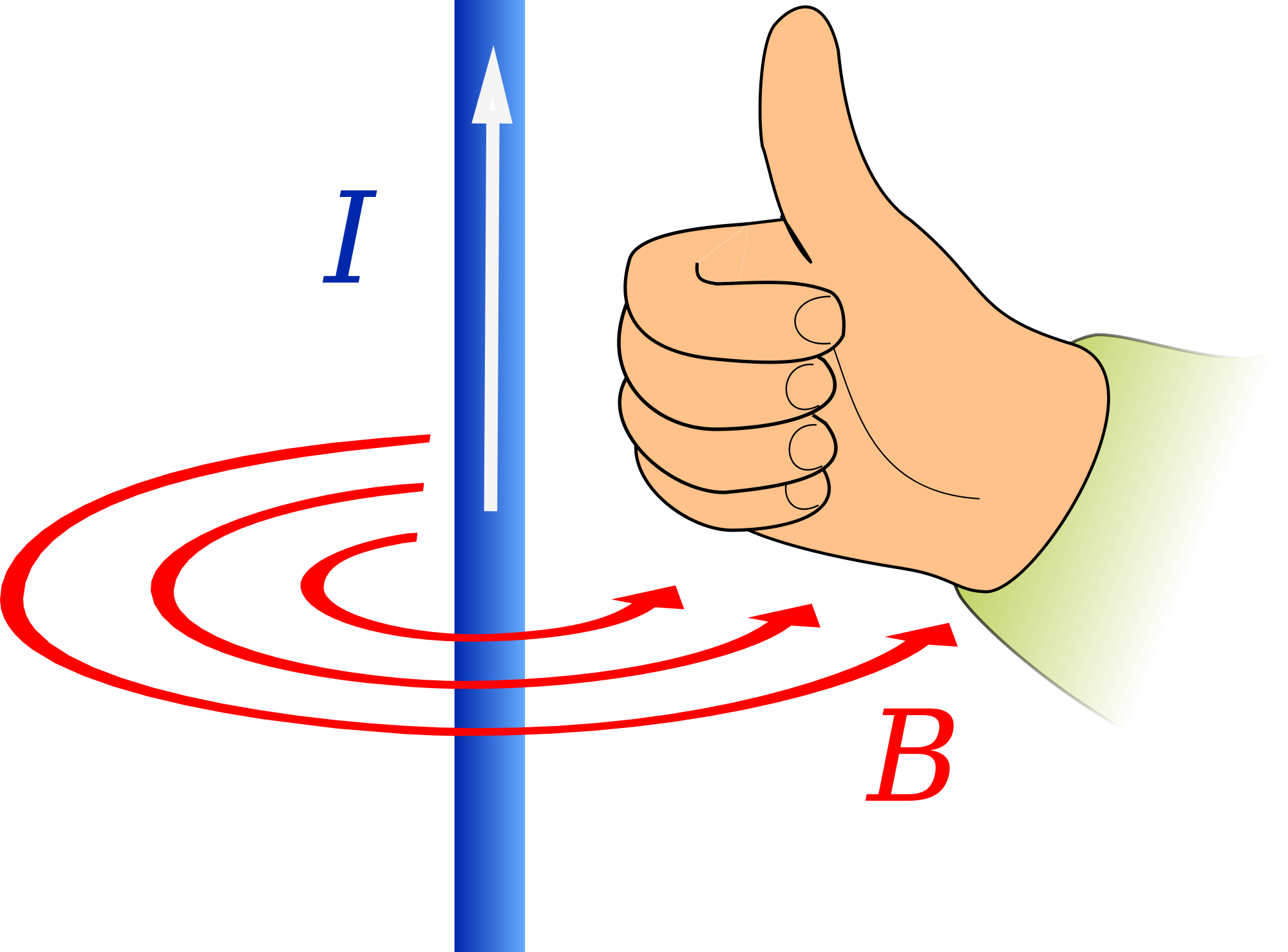}  ~~~~~~~~~~
	\includegraphics[height=1.in]{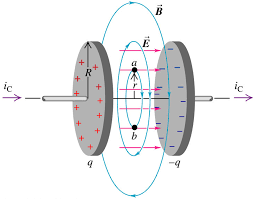}
} 
\centerline{
	(a) ~~~~~~~~~~~~~~~~~~~~~~~~~~~~~~~~~~
	(b) ~~~~~~~~~~~~~~~~~~~~~~~~~~~~~~~~~~
	(c) 
}
\caption{
	(a) Changing magnetic field can generate an electric field around it, that drives
	an electric current in a coil. (b) Electric current $I$ in a wire can generate a
	magnetic field $B$ around it. (c) A changing electric field $E$ (just like
electric current)  can generate a magnetic field $B$ around it.
} 
\label{FAlaw} 
\end{figure}

\begin{figure}[tb] 
\centerline{ 
	\includegraphics[height=1.2in]{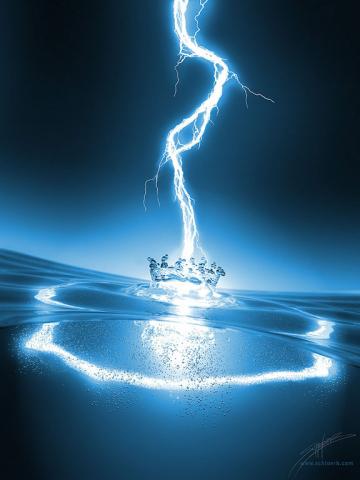}  ~
	\includegraphics[height=1.2in]{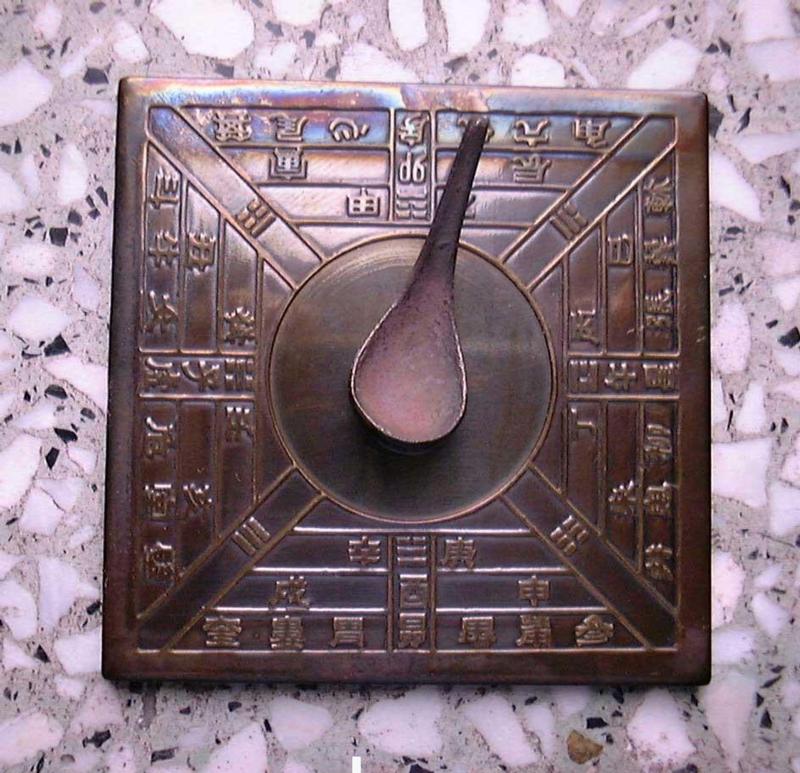}  ~
	\includegraphics[height=1.2in]{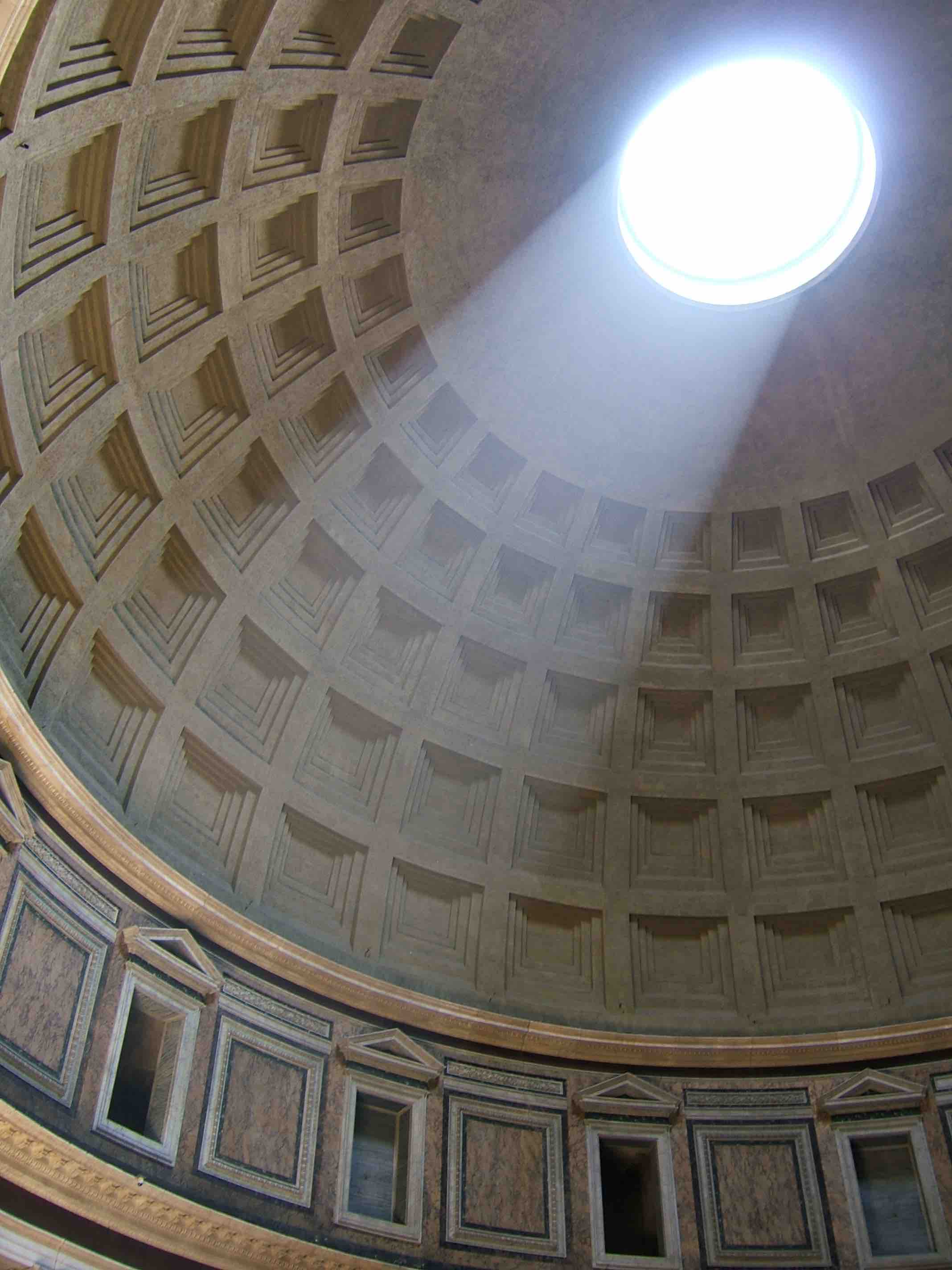} ~~~~~~~
	\includegraphics[height=1.2in]{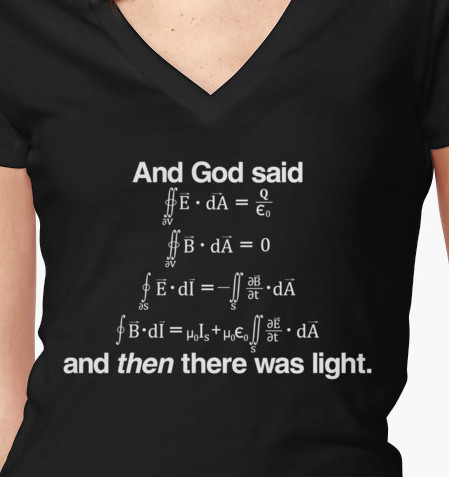}
} 
\caption{
Three very different phenomena, electricity, magnetism, and light,
are unified by Maxwell theory.
} 
\label{eml} 
\end{figure}

But, later we discovered that two other seemingly unrelated phenomena,
electricity and magnetism, can generate each other (see Fig. \ref{FAlaw}).  Our
curiosity about the electricity and magnetism leads to another giant leap in
science, which is summarized by Maxwell equations. Maxwell theory unifies
electricity and magnetism and reveals that light is merely an electromagnetic
wave (see Fig.  \ref{eml}).  We gain a much deeper understanding of light,
which is so familiar and yet so unexpectedly rich and complex in its internal
structure.  This can be viewed as the second revolution -- electromagnetic
revolution.  

\myfrm{
\begin{center}
\textbf{Electromagnetic revolution}
\\
The discovery of a new form of matter -- wave-like matter: electromagnetic
waves, which obey Maxwell equation.  Wave-like matter causes interaction.
\end{center}
}

\begin{figure}[tb] 
\centerline{ 
	\includegraphics[height=1.1in]{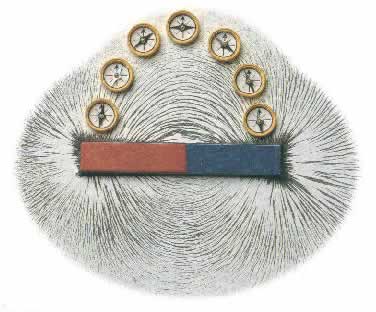}  ~~~~~~~~
	\includegraphics[height=1.1in]{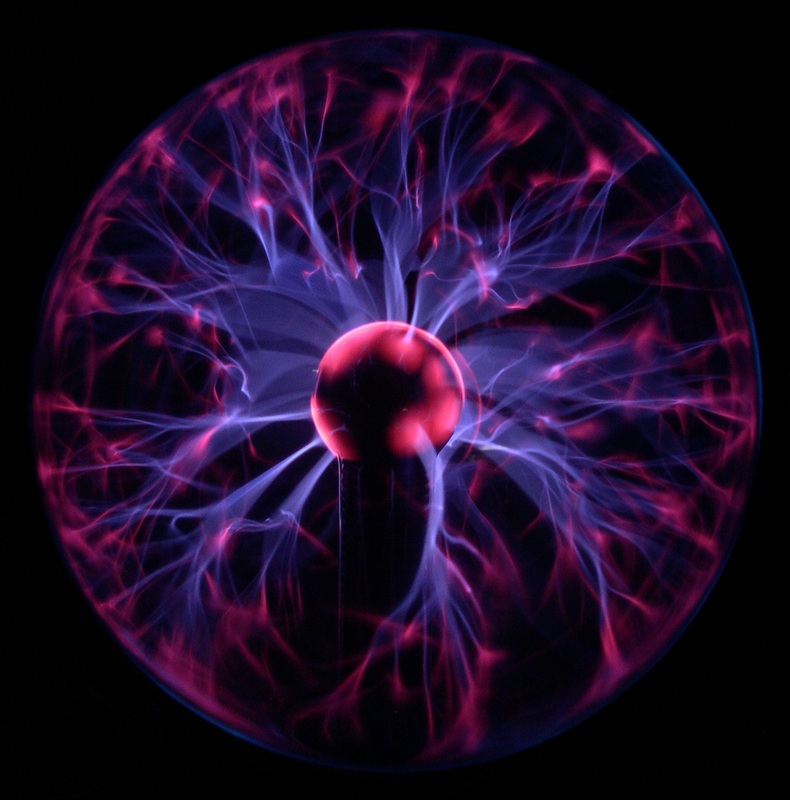}  ~~~~~~~~
	\includegraphics[height=1.3in]{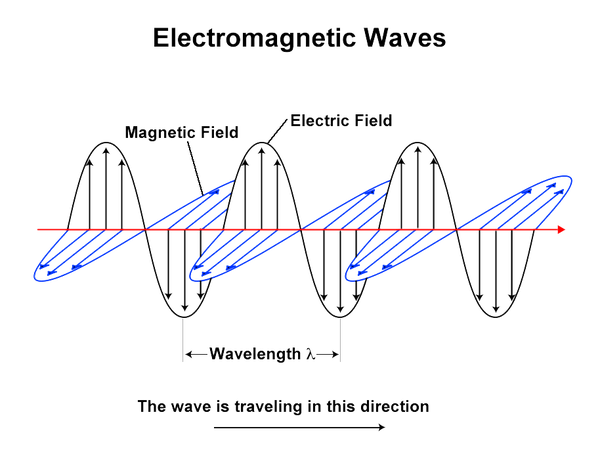}
} 
\centerline{ 
	(a) ~~~~~~~~~~~~~~~~~~~~~~~~~~~~
	(b) ~~~~~~~~~~~~~~~~~~~~~~~~~~~~~~
	(c) ~~
}
\caption{
	(a) Magnetic field revealed by iron powder.
	(b) Electric field revealed by glowing plasma.
	(c) They form a new kind of matter: light -- a wave-like matter
} 
\label{EMfield} 
\end{figure}

However, the true essence of Maxwell theory is the discovery of a new form of
matter -- wave-like (or field-like) matter (see Fig.  \ref{EMfield}), the
electromagnetic wave. The motion of this wave-like matter is governed by
Maxwell equation, which is very different from the particle-like matter
governed by Newton equation $F=ma$.  Thus, the sense that Newton theory
describes everything is incorrect. Newton theory does not apply to wave-like
matter, which requires a new theory -- Maxwell theory.

Unlike the
particle-like matter, the new wave-like matter is closely related to a kind of
interaction -- electromagnetic interaction. In fact, the electromagnetic
interaction can be viewed as an effect of the newly discovered wave-like
matter.

\subsection{Relativity revolution}

After realizing the connection between the interaction and wave-like matter,
one naturally ask: does gravitational interaction also corresponds to a
wave-like matter? The answer is yes.  

\begin{figure}[b] 
\centerline{ 
	\includegraphics[height=1.0in]{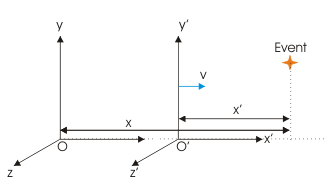}  ~~~~~~~~~~
	\includegraphics[height=0.8in]{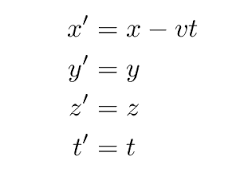}  ~~~~~~~~~~
	\includegraphics[height=1.0in]{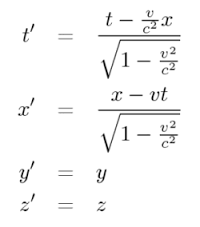}
} 
\centerline{ 
	(a)~~~~~~~~~~~~~~~~~~~~~~~~~~~~~~~~~~~~~~~~
	(b) ~~~~~~~~~~~~~~~~~~~~~~~~~
	(c)
}
\caption{
	(a) A rest frame and a moving frame with velocity $v$.  An event is
recorded with coordinates $(x,y,z,t)$ in the rest frame and
with $(x',y',z',t')$ in the  moving frame.  There are two opinions on how
$(x,y,z,t)$ and $(x',y',z',t')$ are related: (b) Galilean transformation or (c)
Lorantz transformation where $c$ is the speed of light. In our world, the Lorantz transformation is correct.
} 
\label{frame} 
\end{figure}

First, people realized that Newton equation and Maxwell equation have different
symmetries under the transformations between two frames moving against each
other.  In other words,  Newton equation $F=ma$ is invariant under Galileo
transformation, while Maxwell equation is invariant under Lorentz
transformation (see Fig. \ref{frame}).  Certainly, only one of the above two
transformation is correct. If one believes that physical law should be the same
in different frames, then the above observation implies that Newton equation
and Maxwell equation are incompatible, and one of them must be wrong.  If
Galileo transformation is correct, then the Maxwell theory is wrong and needs
to be modified.  If Lorentz transformation is correct, then the Newton theory
is wrong and needs to be modified.  Michelson-Morley experiment showed that the
speed of light is the same in all the frames, which implied the Galileo
transformation to be wrong.  So Einstein choose to believe in Maxwell equation.
He modified Newton equation and developed the theory of special relativity.
Thus, Newton theory is not only incomplete, it is also incorrect.

\begin{figure}[tb] 
\centerline{ 
	\includegraphics[height=1.4in]{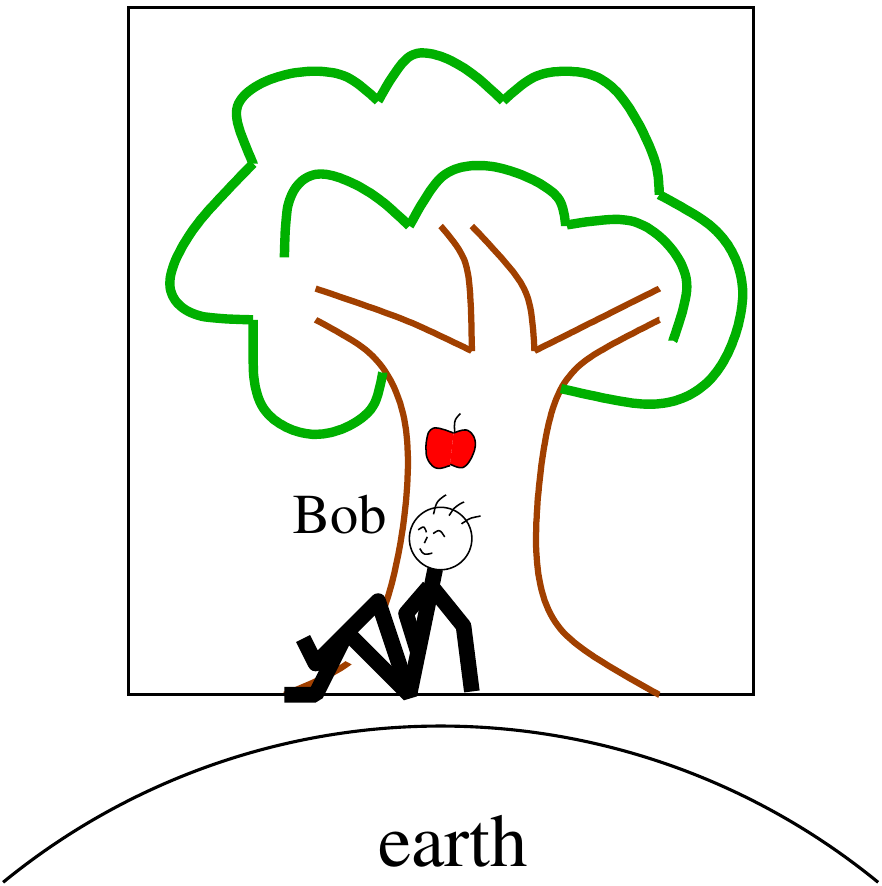}  ~~~
	\includegraphics[height=1.4in]{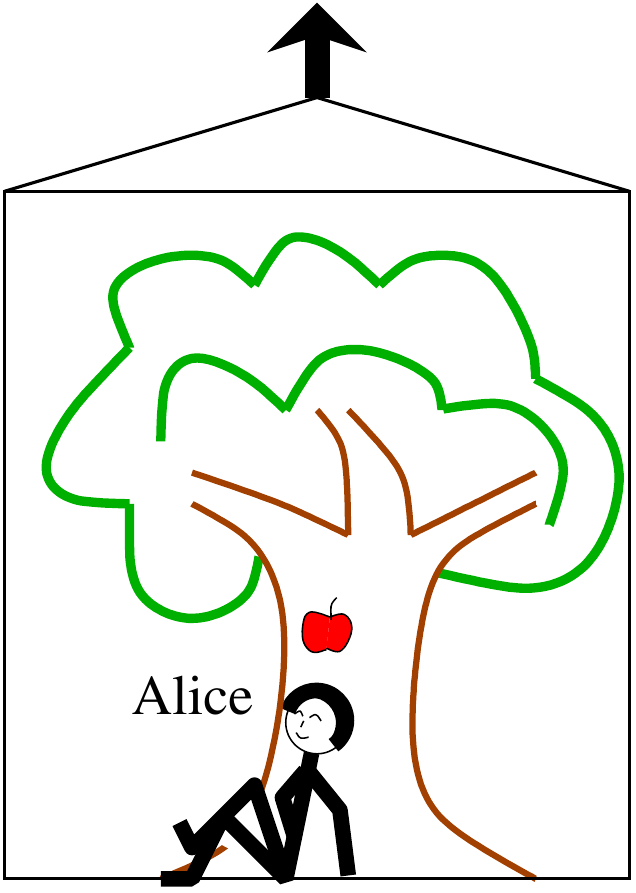}  ~~~~~~~~~~~
	\includegraphics[height=1.2in]{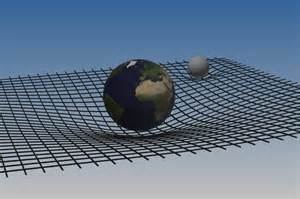}
} 
\caption{
The equivalence of the gravitational force of the earth and the force
experienced in an accelerating elevator, leads to an geometric way to
understand gravity: gravity = distortion in space. In other words, the ``gravitational force'' in an
accelerating elevator is related to a geometric feature: 
the transformation between the
coordinates in a still elevator and in the accelerating elevator.
} 
\label{grav} 
\end{figure}

\begin{figure}[tb] 
\centerline{ 
	\includegraphics[height=1.0in]{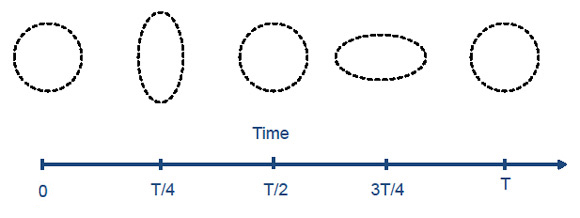}
} 
\caption{
Gravitational wave is a propagating distortion of space:
a circle is distorted by a gravitational wave.
} 
\label{gravwave} 
\end{figure}

Einstein has gone further.  Motivated the equivalence of gravitational force
and the force experienced in an accelerating frame (see Fig. \ref{grav}),
Einstein also developed the theory of general relativity.\cite{E1669} 
Einstein theory unifies several seeming unrelated concepts, such as space and
time, as well as interaction and geometry.  Since the gravity is viewed as a
distortion of space and since the distortion can propagate, Einstein discovered
the second wave-like matter -- gravitational wave (see Fig.  \ref{gravwave}).  
This is another revolution in physics -- relativity
revolution.

\myfrm{
\begin{center}
\textbf{Relativity revolution}
\\
A unification of space and time.
A unification of gravity and space-time distortion.
\end{center}
}

\begin{figure}[tb] 
\centerline{ 
	\includegraphics[height=1.3in]{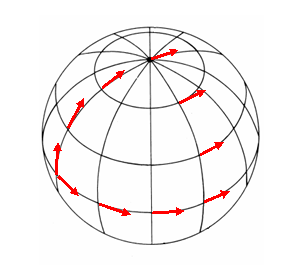}
} 
\caption{
A curved space can be viewed as a distortion of
local directions of the space: parallel moving a local direction (represented by an arrow) around a loop in a curved space,
the direction of the arrow does not come back. Such a twist in  local direction
corresponds to a curvature in space.
} 
\label{ptrans} 
\end{figure}

Motivated by the connection between interaction and geometry in gravity, people
went back to reexamine the electromagnetic interaction, and found that the
electromagnetic interaction is also connected to  geometry.  Einstein's general
relativity views gravity as a distortion of space, which can be viewed as a
distortion of local directions of space (see Fig. \ref{ptrans}).  Motivated by
such a picture, in 1918, Weyl proposed that the unit that we used to measure
physical quantities is relative and is defined only locally.  A distortion of
the unit system can be described by a vector field which is called gauge field.
Weyl proposed that such a vector field (the gauge field) is the vector
potential that describes the electromagnetism.  Although the above particular
proposal turns out to be incorrect, the Weyl's idea is correct.  In 1925, the
complex quantum amplitude was discovered.  If we assume the complex phase is
relative, then a distortion of unit system that measure local complex phase can
also be described by a vector field. Such a vector field is indeed the vector
potential that describes the electromagnetism.  This leads to a unified way to
understand gravity and electromagnetism: gravity arises from the relativity of
spacial directions at different spatial points, while electromagnetism arises
from the relativity of complex quantum phases at different spatial points.
Further more, Nordstr\"om, M\"oglichkeit, Kaluza, and Klein showed that both
gravity and electromagnetism can be understood as a distortion of space-time
provided that we think the space-time as five dimensional with one dimension
compactified into a small circle.\cite{NM1404,K2166,K2695} This can be viewed
as an unification of gravity and electromagnetism.  Those theories are so
beautiful.  Since that time, the geometric way to view our world has dominated
theoretical physics.

\subsection{Quantum revolution}

\begin{figure}[tb] 
\centerline{ 
	\includegraphics[height=1.0in]{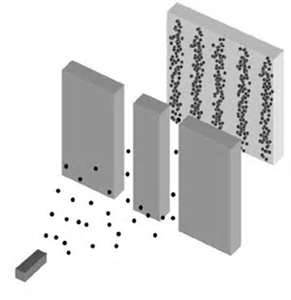} ~~~~
	\includegraphics[height=1.0in]{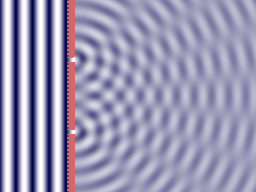} ~~~~~~~~~~~~~~~~~~~~
	\includegraphics[height=1.0in]{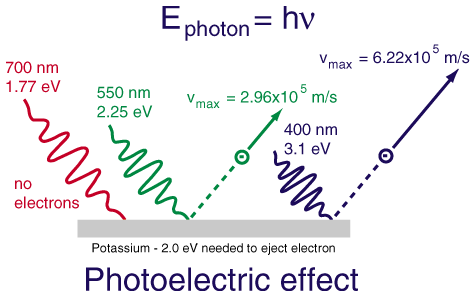}
}
\centerline{ 
	(a) ~~~~~~~~~~~~~~~~~~~~~~~~~~~~~~~~~~~~~~~~~~~~~~~~~~~~~~~~~~~~~~~~~~~~~~~~~~~
	(b)
} 
\caption{
	(a) An electron beam passing through a double-slit can generate an interference
	pattern, indicating that electrons are also waves.  (b) Using light to
eject electrons from a metal (the photoelectric effect) shows that the higher
the light wave frequency (the shorter the wave length), the higher the energy of the ejected electron. This
reveals that a light wave of frequency $f$ can be viewed a beam of particles of
energy $E=h f$, where $h=6.62607004\times 10^{-34} \frac{\text{m}^2\text{kg}}{\text{s}} $.
} 
\label{dslit} 
\end{figure}

However, such a geometric view of world was immediately challenged by new
discoveries from microscopic world.\footnote{Many people have ignored such challenges and the geometric view of world becomes the main stream.}
The experiments in  microscopic world tell us that not
only Newton theory is incorrect, even its relativity modification is incorrect.
This is because Newton theory and its relativistic modification are theories
for particle-like matter.  But through experiments on very tiny things, such as
electrons, people found that the particles are not really particles. They
also behave like waves at the same time.  Similarly, experiments also reveal
that the light waves behave like a beam of particles (photons) at the same time
(see Fig. \ref{dslit}).  So the real matter in our world is not what we thought
it was. The matter is neither particle nor wave, and both particle and wave.
So the Newton theory (and its relativistic modification) for particle-like
matter and the Maxwell/Einstein theories for wave-like matter cannot be the
correct theories for matter. We need a new theory for the new form of
existence: particle-wave-like matter.  The new theory is the quantum theory
that explains the  microscopic world. The quantum theory unifies the
particle-like matter and wave-like matter.  

\myfrm{
\begin{center}
\textbf{Quantum revolution}
\\
There is no particle-like matter nor wave-like matter.
All the matter in our world is particle-wave-like matter.
\end{center}
}

From the above, we see that quantum theory reveals the true existence in our
world to be quite different from the classical notion of existence in our mind.
What exist in our world are not particles or waves, but somethings that are
both particle and wave. Such a picture is beyond our wildest imagination, but
reflects the truth about our world and is the essence of quantum theory.  To
understand the new notion of existence more clearly, let us consider another
example. This time it is about a bit (represented by spin-1/2). A bit has two
possible states of classical existence: $|1\>=|\up \rangle$ and
$|0\>=|\down\rangle $.  However,  quantum theory also allows a new kind of
existence $|\up \rangle + |\down \rangle $.  One may say that $|\up \rangle +
|\down \rangle $ is also a classical existence since $|\up \rangle + |\down
\rangle = |\to \rangle $ that describes a spin in $x$-direction.  So let us
consider a third example of two bits. Then there will be four possible states
of classical existence: $|\up\up \rangle $, $|\up\down \rangle $, $|\down\up
\rangle $, and $|\down\down \rangle $.  Quantum theory allows a new kind of
existence $|\up\up \rangle + |\down\down \rangle $.  Such a quantum existence
is entangled and has no classical analogues.

\begin{figure}[tb] 
\centerline{ 
	\includegraphics[height=.9in]{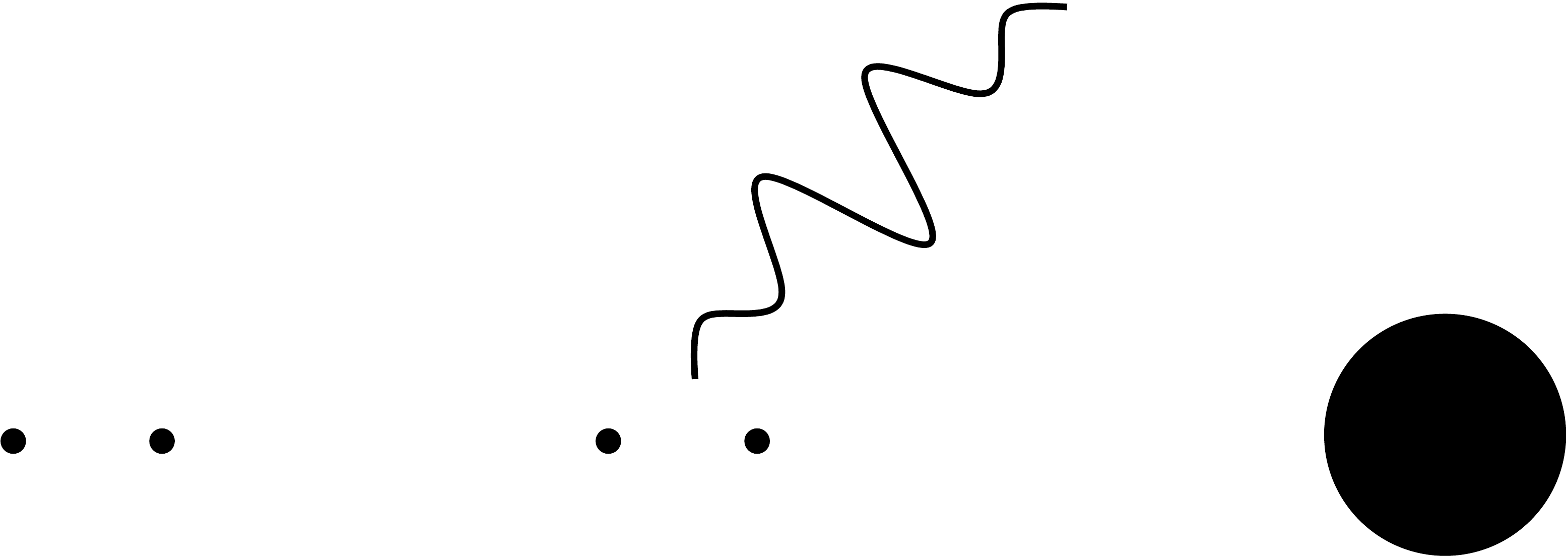}
} 
\caption{
To observe two points of distance $l$ apart, we need to send in light of wave
length $\la < l$.  The corresponding photon has an energy $E=hc/\la$. If $l$ is
less than the Planck length $l<l_P$, then the photon will make a back hole of
size larger then $l$.  The black hole will swallow the two points, and we can
never measure  the separation of two points of distance less than $l_P$.  What
cannot be measured cannot exists.  So the notion of ``two points less than $l_P$
apart'' has no physical meaning and does not exist.
} 
\label{nomani} 
\end{figure}

Although the geometric way to understand our world is a main stream in physics,
here we will take a position that the geometric understanding is not good
enough and will try to advocate a very different non-geometric understanding of
our world.  Why the geometric understanding is not good enough? First the
geometric understanding is not self-consistent. It contradicts with quantum
theory.
The consideration
based quantum mechanics and Einstein gravity indicates that two points
separated by a distance less than the Planck length 
\begin{align}
l_P= \sqrt{\frac{\hbar G}{c^3}} = 1.616199\times 10^{35} \text{m}
\end{align}
cannot exist as a physical reality (see Fig. \ref{nomani}). Thus the foundation
of the geometric approach -- manifold -- simply does not exist in our universe,
since manifold contains points with arbitrary small separation. This suggests
that geometry is an emergent phenomenon that appears only at long distances. So
we cannot use geometry and manifold as a foundation to understand fundamental
physical problems.

Second, Maxwell theory of light and Einstein theory of gravity predict light
waves and gravitational waves.  But the theories fail to tell us what is
waving?  Maxwell theory and Einstein theory are built on top of geometry. They
fail to answer what is the origin of the apparent geometry that we see.  
In other words,  Maxwell theory and Einstein theory are incomplete, and they should be regarded as effective theories at long distances.

Since geometry does not exist in our world, this is why we say the geometric
view of world is challenged by quantum theory.  The quantum theory tell us such
a point of view to be wrong at length scales of order Planck length. So the
quantum theory represents the most dramatic revolution in physics.


\section{It from qubit, not bit  -- A second quantum revolution}


After realizing that even the notion of existence is changed by quantum theory,
it is no longer surprising to see that quantum theory also blurs the
distinction between information and matter.  In fact, it implies that
information is matter, and matter is information \cite{qubit2014}.  This is
because the frequency is an attribute of information.  Quantum theory tells us
that frequency is energy $E=h f$, and relativity tells us that energy is mass
$m=E/c^2$. Both energy and mass are attributes of matter.  So matter =
information. This represents a new way to view our world.

\myfrm{
\begin{center}
\textbf{The essence of quantum theory}
\\
The energy-frequency relation $E=h f$ implies that \textbf{matter = information}.
\end{center}
}

The above point of view of ``matter = information'' is similar to Wheeler's ``it
from bit'', which represents a deep desire to unify matter and information. In
fact, such an unification has happened before at a small scale. We introduced
electric and magnetic field to informationally (or pictorially) describe
electric and magnetic interaction.  But later, electric/magnetic field became
real matter with energy and momentum, and even a particle associated with it.

However, in our world, ``it'' are very complicated. (1) Most ``it'' are
fermions, while ``bit'' are bosonic. Can fermionic ``it'' come from bosonic
``bit''? (2) Most ``it'' also carry spin-1/2. Can spin-1/2 arises from ``bit''?
(3) All ``it'' interact via a special kind of interaction -- gauge interaction.
Can ``bit'' produce gauge interaction? Can ``bit'' produce waves that satisfy
Maxwell equation? Can ``bit'' produce photon?

In other words, to understand the concrete meaning of ``matter from
information'' or ``it from bit'', we note that matter are described by Maxwell
equation (photons), Yang-Mills equation (gluons and $W/Z$ bosons), as well as
Dirac and Weyl equations (electrons, quarks, neutrinos).  The statement
``matter = information'' means that those wave equations can all come from
qubits. In other words, we know that elementary particles (\ie matter) are
described by gauge fields and anti-commuting fields in a quantum field theory.
Here we try to say that all those very different quantum fields can arise from
qubits. Is this possible? 

All the waves and fields mentioned above are waves and fields in space.  The
discovery of gravitational wave strongly suggested that the space is a
deformable dynamical medium. In fact, the discovery of electromagnetic wave and
the Casimir effect already strongly suggested that the space is a deformable
dynamical medium.  As a dynamical medium, it is not surprising that the
deformation of space give rise to various waves.  But the dynamical medium that
describe our space must be very special, since it should give rise to waves
satisfying Einstein equation (gravitational wave), Maxwell equation
(electromagnetic wave), Dirac equation (electron wave), \etc.  But what is the
microscopic structure of the space?  What kind of microscopic structure can, at
the same time, give rise to waves that satisfy Maxwell equation, Dirac/Weyl
equation, and Einstein equation?

Let us view the above questions from another angle.  Modern science has made
many discoveries and has also unified many seemingly unrelated discoveries
into a few simple structures.  Those simple structures are so beautiful and we
regard them as wonders of our universe.  They are also very myterious since we
do not understand where do they come from and why do they have to be the way
they are.  At moment, the most fundamental mysteries and/or wonders in our
universe can be summarized by the following short list:\\
	\textbf{Eight wonders:}\\
	\hspace*{2ex}(1) Locality.\\
	\hspace*{2ex}(2) Identical particles.\\
	\hspace*{2ex}(3) Gauge interactions.\cite{Wey52,P4103,YM5491}\\
	\hspace*{2ex}(4) Fermi statistics.\cite{F2602,D2661}\\
	\hspace*{2ex}(5) Tiny masses of fermions ($\sim 10^{-20}$ of the Planck
mass).\cite{GW7343,P7346,Wqoem}\\
\hspace*{2ex}(6) Chiral fermions.\cite{LY5654,Wo5713,W1301,YX14124784}\\
\hspace*{2ex}(7) Lorentz invariance.\cite{E0591}\\
\hspace*{2ex}(8) Gravity.\cite{E1669}

In the current physical theory of nature (such as the standard model), we take
the above properties for granted and do not ask where do they come from.  We
put those wonderful properties into our theory by hand, for example, by
introducing one field for each kind of interactions or elementary particles.

However, here we would like to question where do those wonderful and mysterious
properties come from?  Following the trend of science history, we wish to have
a single unified understanding of all of the above mysteries.  Or more
precisely, we wish that we can start from a single structure to obtain all of
the above wonderful properties.

The simplest element in quantum theory is qubit $|0\>$ and $|1\>$ (or
$|\down\>$ and $|\up\>$).  Qubit is also the simplest element in quantum
information.  Since our space is a dynamical medium, the simplest choice is to
assume the space to be an ocean of qubits.  We will give such an ocean a
formal name ``qubit ether''.  Then the matter, \ie the elementary particles,
are simply the waves, ``bubbles'' and other defects in the qubit ocean (or
quibt ether).  This is how ``it from qubit'' or ``matter = information''.


Qubit, having only two states $|\down  \rangle $ and $| \up \rangle $, is very
simple.  We may view the many-qubit state with all qubits in $|\down \rangle $
as the quantum state that correspond to the empty space (the vacuum).  Then the
many-qubit state with a few qubits in  $|\up  \rangle $ correspond to a space
with a few spin-0 particles described by a scaler field.  Thus, it is easy to
see that a scaler field can emerge from  qubit ether as a density wave of
up-qubits.  Such a wave satisfy the Eular eqution, but not Maxwell equation or
Yang-Mills equation. So the above particular qubit ether is not the one that
correspond to our space. It has a wrong microscopic structure and cannot carry
waves satisfying Maxwell equation and Yang-Mills equation.  But this line of
thinking may be correct. We just need to find a qubit ether with a different
microscopic structure.

However, for a long time, we do not know how waves satisfying Maxwell equation
or Yang-Mills equation can emerge from any qubit ether.  The anti-commuting
wave that satisfy Dirac/Weyl equation seems even more impossible.  So, even
though quantum theory strongly suggests ``matter = information'', trying to
obtain all elementary particles from an ocean of simple qubits is regarded as
impossible by many and has never become an active research effort.

So the key to understand ``matter = information'' is to identify the microscopic
structure of the qubit ether (which can be viewed as space).  The microscopic
structure of our space must be very rich, since our space
not only can carry gravitational wave and electromagnetic wave, it can also
carry electron wave, quark wave, gluon wave, and the waves that correspond to
all elementary particles. Is such a qubit ether possible?


In condensed matter physics, the discovery of fractional quantum Hall
states\cite{TSG8259} bring us into a new world of highly entangled many-body
systems.  When the strong entanglement becomes long range
entanglement\cite{CGW1038}, the systems will possess a new kind of order --
topological order\cite{Wtop,Wrig}, and represent new states of matter.  We find
that the waves (the excitations) in topologically ordered qubit states can be
very strange: they can be waves that satisfy Maxwell equation, Yang-Mills
equation, or Dirac/Weyl equation.  So the impossible become possible: all
elementary particles (the bosonic force particles and fermionic matter
particles) can emerge from long range entangled qubit ether and be unified by
quantum information
\cite{Wlight,Wqoem,LWqed,GW0600,GW1290,W1281,W1301,YX14124784,ZW150802595}.

We would like to stress that the above picture is ``it from qubit'', which is
very different from Wheeler's ``it from bit''. As we have explained, our
observed elementary particles can only emerge from  long range entangled qubit
ether.  The requirement of quantum entanglement implies that ``it cannot from
bit''.  In fact ``it from entangled qubits''.

\section{A string-net liquid of qubits and a
unification of gauge interactions and Fermi statistics}

In this section, we will consider a particular entangled qubit ocean -- a
string liquid of qubits.  Such entangled qubit ocean support new kind of waves
and their corresponding particles.  We find that the new waves and the emergent
statistics are so profound, that they may change our view of universe. Let us
start by explaining a basic notion -- ``principle of emergence''.

\subsection{Principle of emergence} \label{pem}

Typically, one thinks the properties of a material should be determined by the
components that form the material.  However, this simple intuition is
incorrect, since all the materials are made of same components: electrons,
protons and neutrons.  So we cannot use the richness of the components to
understand the richness of the materials. In fact, the various properties of
different materials originate from various ways in which the particles are
organized.  Different orders (the organizations of particles) give rise to
different physical properties of a material. It is the richness of the orders
that gives rise to the richness of material world.


\begin{figure}[t]
\centerline{
\includegraphics[scale=0.8]{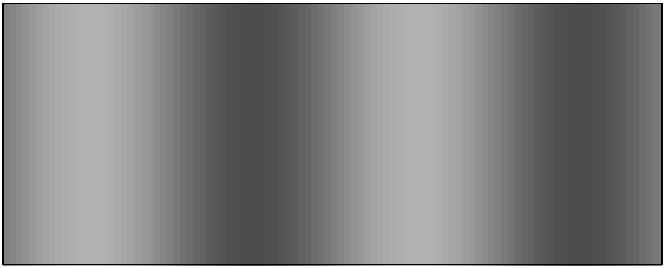}
}
\caption{
Liquids only have a compression wave -- a wave of density fluctuations.
}
\label{liquidC}
\end{figure}

Let us use the origin of mechanical properties and the origin of waves to
explain, in a more concrete way, how orders determine the physics properties of
a material.  We know that a deformation in a material can propagate just like
the ripple on the surface of water.  The propagating deformation corresponds to
a wave traveling through the material.  Since liquids can resist only
compression deformation, so liquids can only support a single kind of wave --
compression wave (see Fig.  \ref{liquidC}). (Compression wave is also
called longitudinal wave.) Mathematically the motion of the compression wave is
governed by the Euler equation
\begin{equation}
\label{EulEq}
 \frac{\prt^2 \rho}{\prt t^2}-v^2 \frac{\prt^2 \rho}{\prt x^2}=0,
\end{equation}
where $\rho$ is the density of the liquid.  

Solid can resist both compression and shear deformations.  As a result, solids
can support both compression wave and transverse wave. The transverse wave
correspond to the propagation of shear deformations. In fact there are two
transverse waves corresponding to two directions of shear deformations.  The
propagation of the compression wave and the two transverse waves in solids are
described by the elasticity equation
\begin{equation}
\label{NavEq}
 \frac{\prt^2 u^i}{\prt t^2}- T^{ikl}_j \frac{\prt^2 u^j}{\prt x^k\prt x^l}
=0
\end{equation}
where the vector field $u^i(\v x, t)$  describes the local displacement of the
solid. 

\begin{figure}[t]
\centerline{
\includegraphics[scale=0.42]{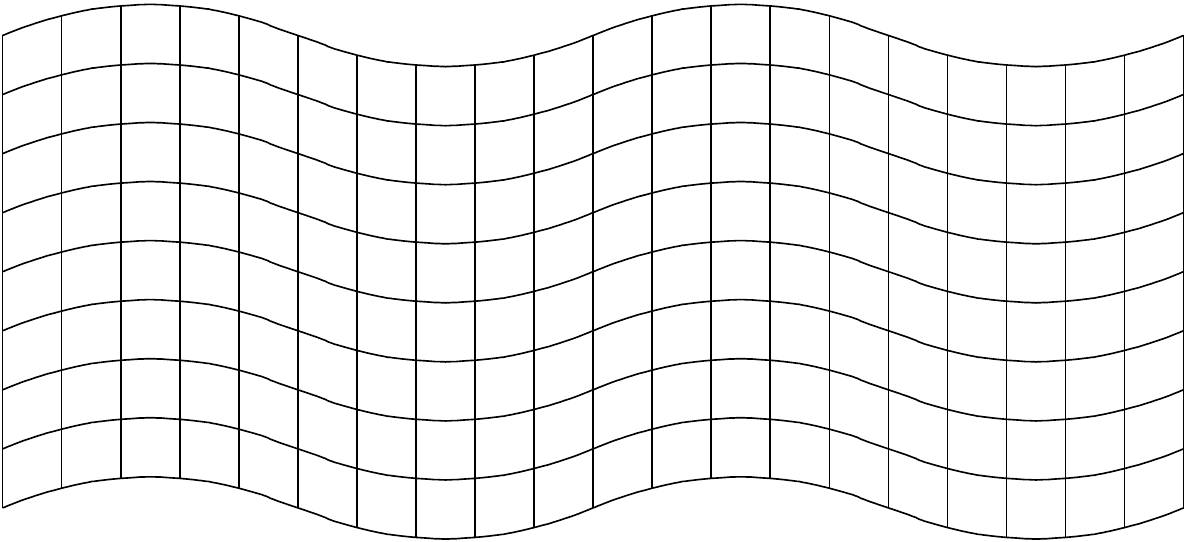}
}
\caption{
Drawing a grid on a sold helps us to see the deformation of the solid.  The
vector $u^i$ in \eqref{NavEq} is the displacement of a vertex in the grid.  In
addition to the compression wave (\ie the density wave), a solid also supports
transverse wave (wave of shear deformation) as shown in the above figure.
}
\label{crystalC}
\end{figure}

We would like to point out that the elasticity equation and the Euler equations
not only describe the propagation of waves, they actually describe all small
deformations in solids and liquids.  Thus, the two equations represent a
complete mathematical description of the mechanical properties of solids and
liquids.  

\begin{figure}[t]
\centerline{
\includegraphics[scale=0.5]{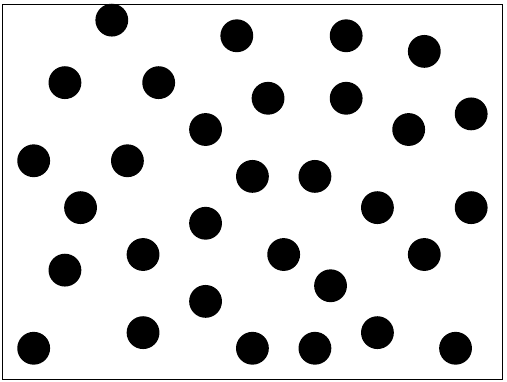}
\hfil
\includegraphics[scale=0.5]{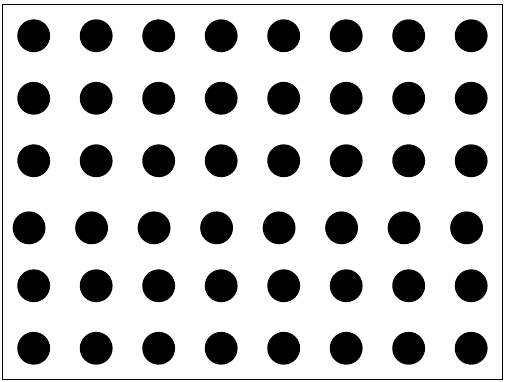}
}
\centerline{
\hfil
(a)
\hfil\hfil
(b)
\hfil
}
\caption{
(a) Particles in liquids do not have fixed relative positions.
They fluctuate freely and have a random but uniform distribution.
(b) Particles in solids form a fixed regular lattice.
}
\label{cryliq1}
\end{figure}

But why do solids and liquids behave so differently? What makes a solid to have
a shape and a liquid to have no shape?  What are the origins of
elasticity  equation and Euler equations? The answer to those questions has to
wait until the discovery of atoms in 19th century.  Since then, we realized
that both solids and liquids are formed by collections of atoms. The main
difference between the solids and liquids is that the atoms are organized very
differently.  In liquids, the positions of atoms fluctuate randomly (see Fig.
\ref{cryliq1}a),  while in solids, atoms organize into a regular fixed array
(see Fig.  \ref{cryliq1}b).\footnote{The solids here should be more accurately
referred as crystals.}  It is the different organizations of atoms that lead to
the different mechanical properties of liquids and solids.  In other words, it
is the different organizations of atoms that make liquids to be able to flow
freely and solids to be able to retain its shape.

\begin{figure}[t]
\centerline{
\includegraphics[scale=0.5]{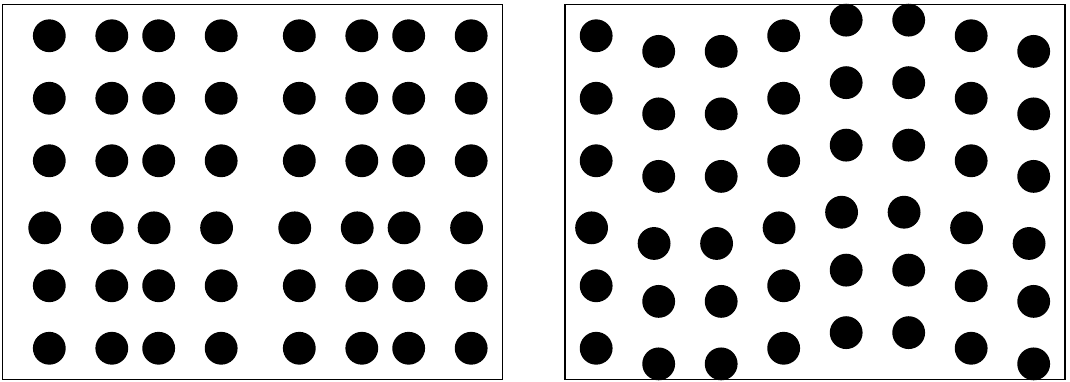}
}
\centerline{
(a)  ~~~~~~~~~~~~~~~~~
(b)
}
\caption{
The atomic picture of 
(a) the compression wave and
(b) the transverse wave in a crystal.
}
\label{sndwavs1}
\end{figure}

\begin{figure}[t]
\centerline{
\includegraphics[scale=0.5]{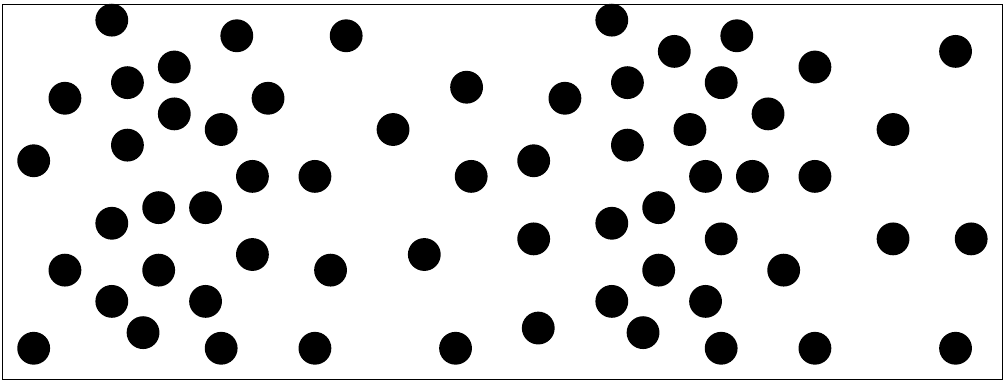}
}
\caption{
The atomic picture of the compression wave in liquids. 
}
\label{liquidwav1}
\end{figure}

How can different organizations of atoms affect mechanical properties of
materials?  In solids, both the compression deformation (see Fig.
\ref{sndwavs1}a) and the shear deformation (see Fig. \ref{sndwavs1}b) lead to
real physical changes of the atomic configurations. Such changes cost energies.
As a result, solids can resist both kinds of deformations and can retain their
shapes.  This is why we have both the compression wave and the transverse wave
in solids.

In contrast, a shear deformation of atoms in liquids does not result in a new
configuration since the atoms still have uniformly random positions. So the
shear deformation is a do-nothing operation for liquids.  Only the compression
deformation which changes the density of the atoms results in a new atomic
configuration and costs energies. As a result, liquids can only resist
compression and have only compression wave.  Since shear deformations do not
cost any energy for liquids, liquids can flow freely.

We see that the properties of the propagating wave are entirely determined by
how the atoms are organized in the materials.  Different organizations lead to
different kinds of waves and different kinds of mechanical laws.  Such a point
of view of different kinds of waves/laws originated from different
organizations of 
particles is a central theme in condensed matter physics.  This point of view
is called the principle of emergence.  

\subsection{String-net liquid of qubits unifies light and electrons}

The elasticity equation and the Euler equation are two very important
equations.  They lay the foundation of many branches of science such as
mechanical engineering, aerodynamic engineering, \etc.  But, we have a more
important equation, Maxwell equation, that describes light waves in vacuum.
When Maxwell equation was first introduced, people firmly believed that any
wave must corresponds to motion of something.  So people want to find out
what is the origin of the Maxwell equation?  The motion of what gives rise
electromagnetic wave?

First, one may wonder: can Maxwell equation comes from a certain symmetry
breaking order?  Based on Landau symmetry-breaking theory, the different
symmetry breaking orders can indeed lead to different waves satisfying
different wave equations.
So maybe a certain symmetry breaking order can give rise to a wave that satisfy
Maxwell equation.  But people have been searching for ether -- a medium that
supports light wave -- for over 100 years, and could not find any symmetry
breaking states that can give rise to waves satisfying the Maxwell equation.
This is one of the reasons why people give up the idea of ether as the origin
of light and Maxwell equation.

However, the discovery of topological order\cite{Wtop,Wrig} suggests that
Landau symmetry-breaking theory does not describe all possible organizations of
bosons/spins.  This gives us a new hope: Maxwell equation may arise from a new
kind of organizations of bosons/spins that have non-trivial topological orders.

In addition to the Maxwell equation, there is an even stranger equation, Dirac
equation, that describes wave of electrons (and other fermions).  Electrons
have Fermi statistics. They are fundamentally different from the quanta of
other familiar waves, such as photons and phonons, since those quanta all have
Bose statistics.  To describe the electron wave, the amplitude of the wave must
be anti-commuting Grassmann numbers, so that the wave quanta will have Fermi
statistics. Since electrons are so strange, few people regard electrons and the
electron waves as collective motions of something. People accept without
questioning that electrons are fundamental particles, one of the building
blocks of all that exist.  

However, from a condensed matter physics point of view, all low energy
excitations are collective motion of something.  If we try to regard photons as
collective modes, why cann't we regard electrons as collective modes as well?
So maybe, Dirac equation and the associated fermions can also arise from a new
kind of organizations of bosons/spins that have non-trivial topological orders.

\begin{figure}[tb]
\centerline{
\includegraphics[scale=0.13]{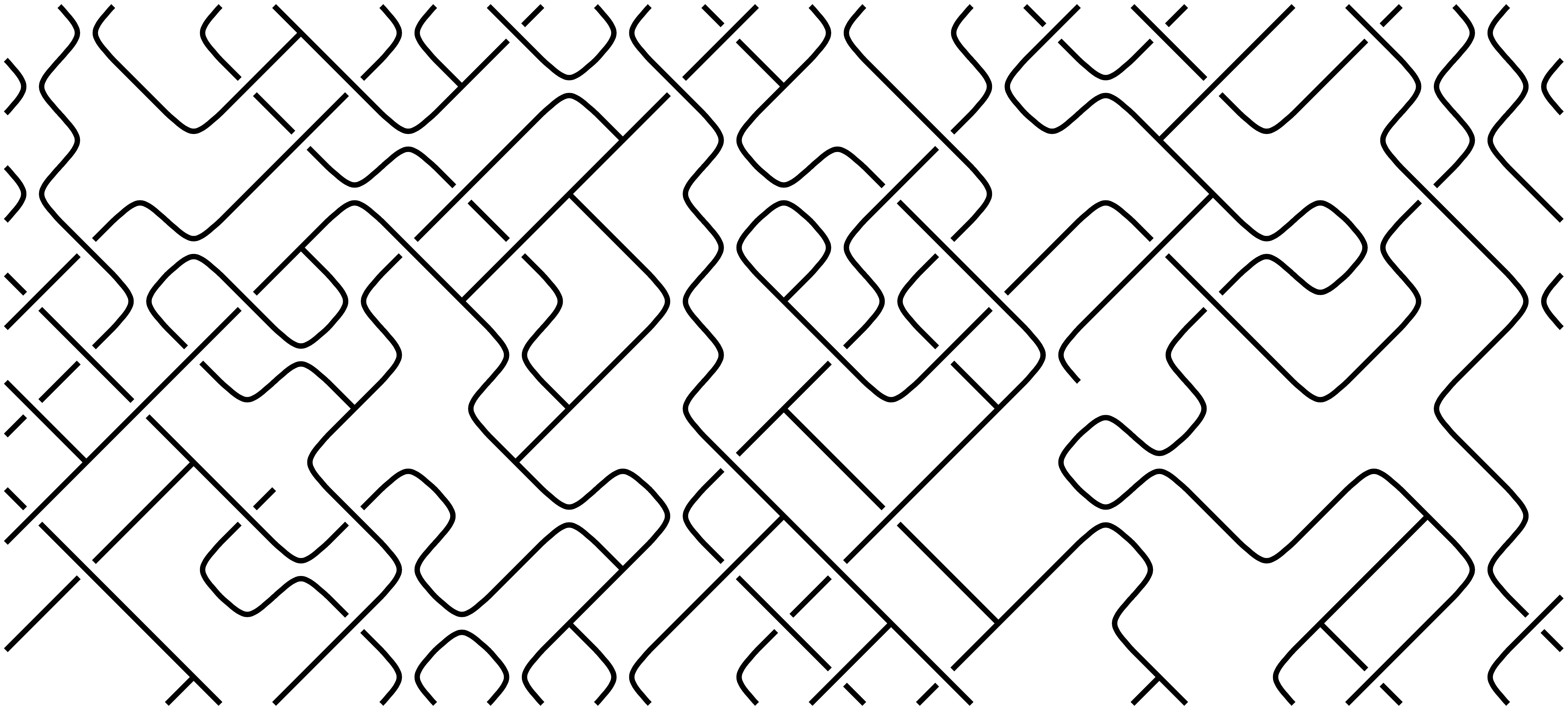}
}
\caption{
A quantum ether: The fluctuation of oriented strings give rise to
electromagnetic waves (or light). The ends of strings give rise to electrons.
Note that oriented strings have directions which should be described by curves
with arrow. For ease of drawing, the arrows on the curves are omitted in the
above plot.
}
\label{stringnetS}
\end{figure}

A recent study provides an positive answer to the above
questions.\cite{LWstrnet,LWuni,LWqed} We find that if bosons/spins form large
oriented strings and if those strings form a quantum liquid state, then the
collective motion of the such organized bosons/spins will correspond to waves
described by Maxwell equation and Dirac equation.  The strings in the string
liquid are free to join and cross each other. As a result, the strings look
more like a network (see Fig.  \ref{stringnetS}).  For this reason, the string
liquid is actually a liquid of string-nets, which is called string-net
condensed state.

\begin{figure}[tb]
\centerline{
\includegraphics[width=2.5in]{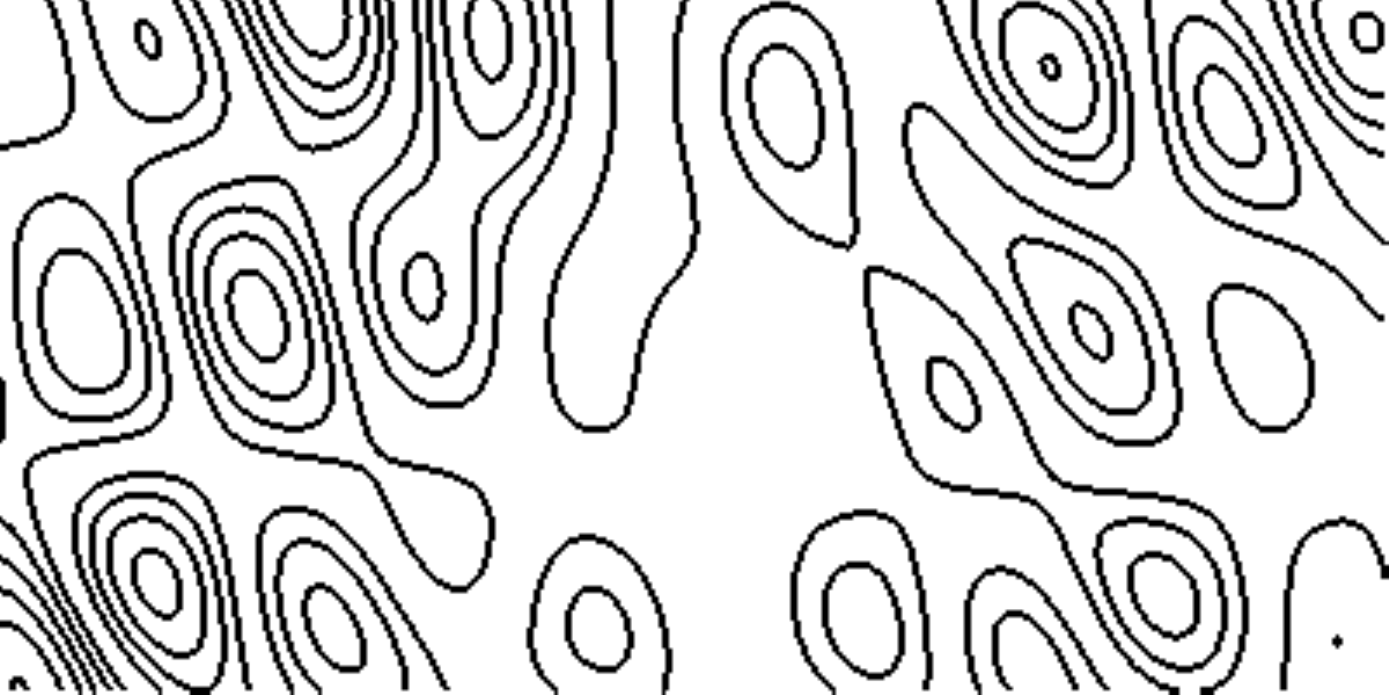}
}
\caption{
The fluctuating strings in a string liquid.
}
\label{vacBW}
\end{figure}

\begin{figure}[tb]
\centerline{
\includegraphics[width=2.5in]{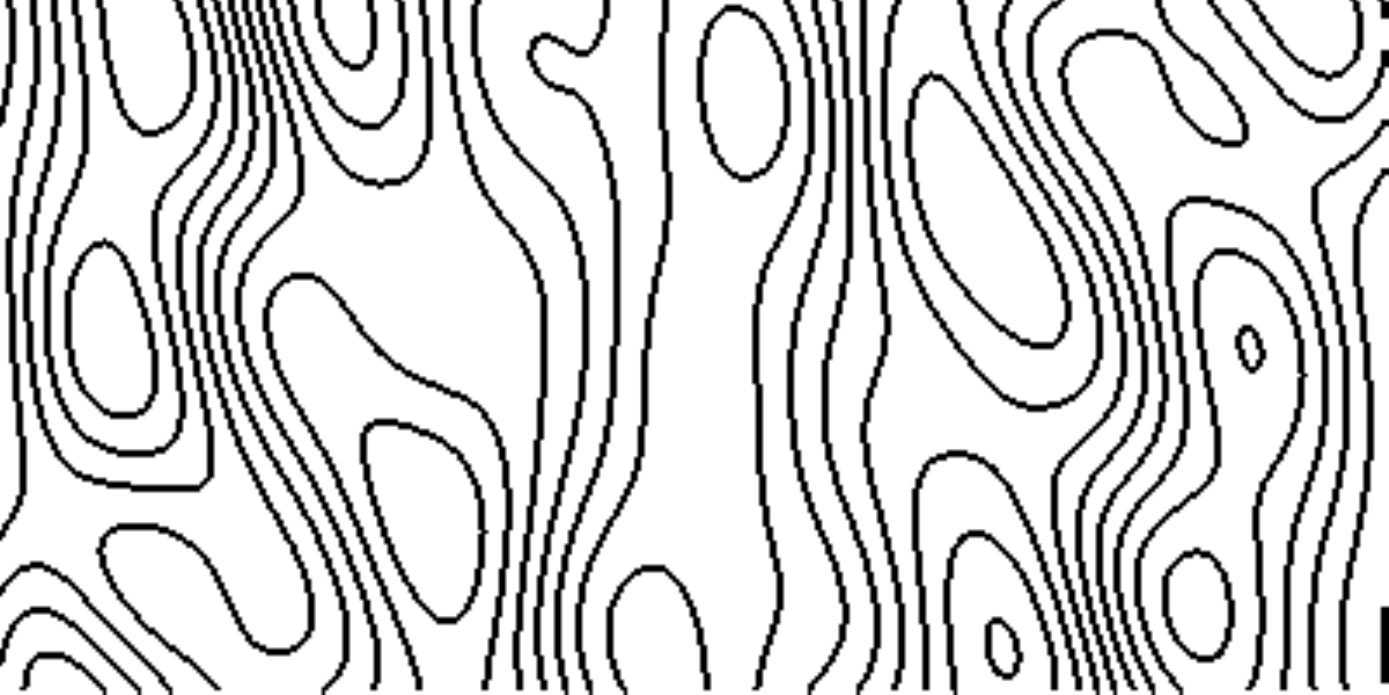}
}
\caption{
A ``density'' wave of oriented strings in a string liquid.  The wave propagates
in $\v x$-direction.  The ``density'' vector $\v E$ points in $\v y$-direction.
For ease of drawing, the arrows on the oriented strings are omitted in the
above plot.
}
\label{vacwvBW}
\end{figure}

But why the waving of strings produces waves described by the Maxwell equation?
We know that the particles in a liquid have a random but uniform distribution.
A deformation of such a distribution corresponds a density fluctuation, which
can be described by a scaler field $\rho(\v x,t)$.  Thus the waves in a liquid
is described by the scaler field $\rho(\v x,t)$ which satisfy the Euler
equation \eqref{EulEq}.  Similarly, the strings in a string-net liquid also have
a random but uniform distribution (see Fig.  \ref{vacBW}). A deformation of
string-net liquid corresponds to a change of the density of the strings (see
Fig. \ref{vacwvBW}).  However, since strings have an orientation, the
``density'' fluctuations are described by a vector field $\v E(\v x, t)$, which
indicates there are more strings in the $\v E$ direction on average.  The
oriented strings can be regarded as flux lines. The vector field $\v E(\v x,
t)$ describes the smeared average flux.  
Since strings are continuous (\ie they cannot end), the
flux is conserved: $\v\prt\cdot\v E(\v x, t)=0$.  The vector density $\v E(\v
x, t)$ of strings cannot change in the direction along the strings (\ie along
the $\v E(\v x, t)$ direction). $\v E(\v x, t)$ can change only in the
direction perpendicular to $\v E(\v x, t)$.  Since the direction of the
propagation is the same as the direction in which $\v E(\v x, t)$ varies, thus
the waves described by $\v E(\v x, t)$ must be transverse waves: $\v E(\v x,
t)$ is always perpendicular to the direction of the propagation.  Therefore,
the waves in the string liquid have a very special property: the waves have
only transverse modes and no longitudinal mode.  This is exactly the property
of the light waves described by the Maxwell equation.  We see that ``density''
fluctuations of strings (which are described be a transverse vector field)
naturally give rise to the light (or electromagnetic) waves and the Maxwell
equation.\cite{Walight,Wqoem,MS0312,HFB0404,LWuni,LWqed}

It is interesting to compare solid, liquid, and string-net liquid.
We know that the particles in a solid organized into a regular
lattice pattern.  The waving of such organized particles produces a compression
wave and two transverse waves.  The particles in a liquid have a more random
organization.  As a result, the waves in liquids lost two transverse modes and
contain only a single compression mode.  The particles in a string-net liquid
also have a random organization, but in a different way.  The particles first
form string-nets and string-nets then form a random liquid state. Due to this
different kind of randomness, the waves in string-net condensed state lost the
compression mode and contain two transverse modes.  Such a wave (having only
two transverse modes) is exactly the electromagnetic wave.

To understand how electrons appear from string-nets, we would like to point out
that if we only want photons and no other particles, the strings must be closed
strings with no ends.  The fluctuations of closed strings produce only photons.
If strings have open ends, those open ends can move around and just behave like
independent particles.  Those particles are not photons. In fact, the ends of
strings are nothing but electrons.

How do we know that ends of strings behave like electrons?  First, since the
waving of string-nets is an electromagnetic wave, a deformation of string-nets
correspond to an electromagnetic field.  So we can study how an end of a string
interacts with a deformation of string-nets.  We find that such an interaction
is just like the interaction between a charged electron and an electromagnetic
field. Also electrons have a subtle but very important property -- Fermi
statistics, which is a property that exists only in quantum theory.  
Amazingly, the ends of strings can reproduce this subtle quantum property of
Fermi statistics.\cite{LWsta,LWstrnet}  Actually, string-net liquids explain
why Fermi statistics should exist.

We see that qubits that organize into string-net liquid naturally explain both
light and electrons (gauge interactions and Fermi statistics).  In other words,
string-net theory provides a way to unify light and
electrons.\cite{LWuni,LWqed} So, the fact that our vacuum contains both light
and electrons may not be a mere accident. It may actually suggest that the
vacuum is indeed a string-net liquid.  

\subsection{More general string-net liquid and emergence of Yang-Mills gauge
theory}

Here, we would like to point out that there are many different kinds of
string-net liquids.  
The strings in different liquids may have different numbers of types.  The
strings may also join in different ways.  For a general string-net liquid, the
waving of the strings may not correspond to light and the ends of strings may
not be electrons.  Only one kind of string-net liquids give rise to light and
electrons.  On the other hand, the fact that there are many different kinds of
string-net liquids allows us to explain more than just light and electrons.  We
can design a particular type of string-net liquids which not only gives rise to
electrons and photons, but also gives rise to quarks and
gluons.\cite{Wqoem,LWstrnet} The waving of such type of string-nets corresponds
to photons (light) and gluons. The ends of different types of strings
correspond to electrons and quarks. It would be interesting to see if it is
possible to design a string-net liquid that produces all elementary particles!
If this is possible, the ether formed by such string-nets can provide an origin
of all elementary particles.\footnote{So far we can use string-net to produce
almost all elementary particles, expect for the graviton that is responsible
for the gravity.  In particular, we can even produce the chiral coupling between
the $SU(2)$ gauge boson and the fermions from the qubit ocean \cite{W1301,YX14124784}.}

We like to stress that the string-nets are formed by qubits.  So in the
string-net picture, both the Maxwell equation and Dirac equation, emerge from
\emph{local} qubit model, as long as the qubits from a long-range entangled
state (\ie a string-net liquid).  In other words, light and electrons are
unified by the long-range entanglement of qubits!

The electric field and the magnetic field in the Maxwell equation are called
gauge fields.  The field in the Dirac equation are Grassmann-number valued
field.\footnote{Grassmann numbers are anti-commuting numbers.} For a long time,
we thought that we have to use gauge fields to describe light waves that have
only two transverse modes, and we thought that we have to use Grassmann-number
valued fields to describe electrons and quarks that have Fermi statistics. So
gauge fields and Grassmann-number valued fields become the fundamental build
blocks of quantum field theory that describe our world.  The string-net liquids
demonstrate we do not have to introduce  gauge fields and  Grassmann-number
valued fields to describe photons, gluons, electrons, and quarks. It
demonstrates how gauge fields and Grassmann fields emerge from local qubit
models that contain only complex scaler fields at the cut-off scale.  

Our attempt to understand light has long and evolving history.  We first
thought light to be a beam of particles.  After Maxwell, we understand light as
electromagnetic waves.  After Einstein's theory of general relativity, where
gravity is viewed as curvature in space-time, Weyl and others try to view
electromagnetic field as curvatures in the ``unit system'' that we used to
measure complex phases.  It leads to the notion of gauge theory.  The general
relativity and the gauge theory are two corner stones of modern physics. They
provide a unified understanding of all four interactions in terms of a
beautiful mathematical frame work: all interactions can be understood
geometrically as curvatures in space-time and in ``unit systems'' (or more
precisely, as curvatures in the tangent bundle and other vector bundles in
space-time).

Later, people in high-energy physics and in condensed matter physics have found
another way in which gauge field can emerge:\cite{DDL7863,W7985,BA8880,AM8874}
one first cut a particle (such as an electron) into two partons by writing the
field of the particle as the product of the two fields of the two partons.
Then one introduces a gauge field to glue the  two partons back to the original
particle.  Such a ``glue-picture'' of gauge fields (instead of the fiber
bundle picture  of gauge fields) allow us to understand the emergence of gauge
fields in models that originally contain no gauge field at the cut-off scale.

\begin{figure}[tb]
\centerline{
	\includegraphics[height=0.9in]{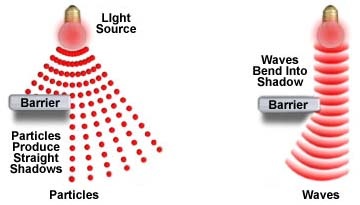}~~~~~
	\includegraphics[height=1.1in]{EMwave}~~~~~
	\includegraphics[height=0.9in]{ptrans}
}
\centerline{
(a) ~~~~~~~~~~~~~~~~~
(b) ~~~~~~~~~~~~~~~~~~~
(c) ~~~~~~~~~~~~~~~~~~~~
(d) ~~
}
\centerline{
	\includegraphics[height=1.4in]{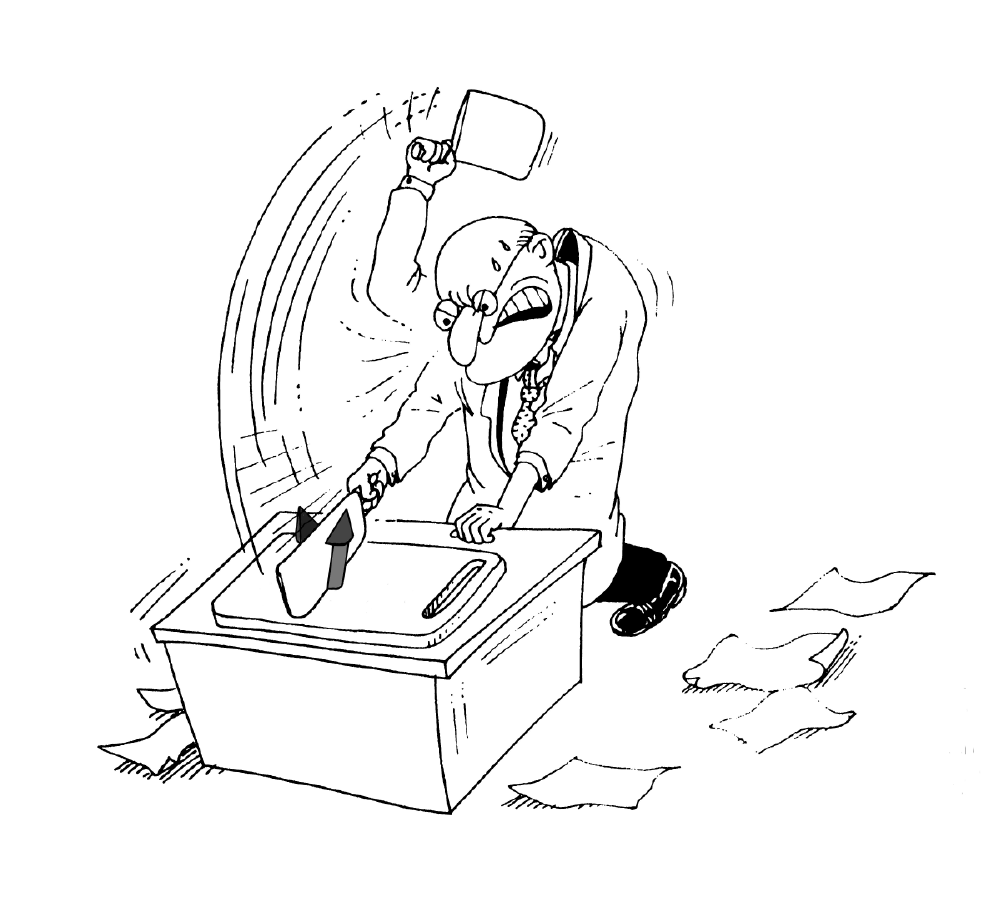}~~~~~
	\includegraphics[height=0.75in]{vacwvBW}~~~~~~~~~~~~
	\includegraphics[height=1.0in]{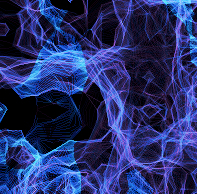}
}
\centerline{
(e) ~~~~~~~~~~~~~~~~~~~~~~~~~~~~~~~~~
(f) ~~~~~~~~~~~~~~~~~~~~~~~~~~~~~~~~~~
(g)
}
\caption{
The evolution of our understanding of light (and gauge interaction): 
(a) particle beam, (b) wave, (c) electromagnetic wave, (d)
curvature in fiber bundle, (e) glue of partons, (f) wave in string-net liquid,
(g)  wave in long-range entanglement of many qubits.
}
\label{gauge}
\end{figure}

A string picture represent the third way to understand gauge theory.  String
operators appear in the Wilson-loop characterization\cite{W7445} of gauge
theory. The Hamiltonian and the duality description of lattice gauge theory
also reveal string structures.\cite{KS7595,BMK7793,K7959,S8053}.  Lattice gauge
theories are not local bosonic models and the strings are unbreakable in
lattice gauge theories.  String-net theory points out that even breakable
strings can give rise to gauge fields.\cite{HWcnt} So we do not really need
strings. Qubits themselves are capable of generating gauge fields
and the associated Maxwell equation.  This phenomenon was discovered in several
qubit models\cite{FNN8035,BA8880,Wlight,MS0204,HFB0404} before realizing
their connection to the string-net liquids.\cite{Walight}  Since gauge field
can emerge from local qubit models, the string picture
evolves into the entanglement picture -- the fourth way to understand gauge
field: gauge fields are fluctuations of long-range entanglement.  I feel that
the  entanglement picture capture the essence of gauge theory.  Despite the
beauty of the geometric picture, the essence of gauge theory is not the curved
fiber bundles.  In fact, we can view gauge theory as a theory for long-range
entanglement, although the gauge theory is discovered long before the notion
of long-range entanglement. The evolution of our understanding of light and
gauge interaction: particle beam $\to$ wave $\to$ electromagnetic wave $\to$
curvature in fiber bundle $\to$ glue of partons $\to$ wave in string-net liquid
$\to$  wave in long-range entanglement, represents 200 year's effort of human
race to unveil the mystery of universe (see Fig. \ref{gauge}).

Viewing gauge field (and the associated gauge bosons) as fluctuations of
long-range entanglement has an added bonus: we can understand the origin of
Fermi statistics in the same way: fermions emerge as defects of long-range
entanglement, even though the original model is purely bosonic.  Previously,
there are two ways to obtain emergent fermions from purely bosonic model: by
binding gauge charge and gauge flux in (2+1)D,\cite{LM7701,W8257} and by
binding the charge and the monopole in a $U(1)$ gauge theory in
(3+1)D.\cite{T3141,JR7616,W8246,G8205,LM0012} But those approaches only work in
(2+1)D or only for $U(1)$ gauge field.  Using long-range entanglement and their
string-net realization, we can obtain the simultaneous emergence of both gauge
bosons and fermions in \emph{any} dimensions and for any gauge
group.\cite{LWsta,LWstrnet,LWuni,Wqoem} This result gives us hope that maybe
every elementary particles are emergent and can be unified using local qubit
models.  Thus, long-range entanglement offer us a new option to view our world:
maybe our vacuum is a long-range entangled state.  It is the pattern of the
long-range entanglement in the vacuum that determines the content and the
structures of observed elementary particles.  Such a picture has an
experimental prediction that will be described in the next section \ref{fPred}.

We like to point out that the string-net unification of gauge
bosons and fermions is very different from the superstring theory for gauge
bosons and fermions.  In the string-net theory, gauge bosons and fermions come
from the qubits that form the space, and ``string-net'' is simply the name that
describe how qubits are organized in the ground state.  So string-net is not a
thing, but a pattern of qubits.  In the string-net theory, the gauge bosons are
waves of collective fluctuations of the string-nets, and a fermion corresponds
to one end of string.  In contrast, gauge bosons and fermions come from strings
in the superstring theory. Both gauge bosons and fermions correspond to small
pieces of strings.  Different vibrations of the small pieces of strings give
rise to different kind of particles. The fermions in the superstring theory are
put in by hand through the introduction of Grassmann fields.

\subsection{A falsifiable prediction of 
string-net unification of gauge interactions and Fermi statistics}
\label{fPred}

In the string-net unification of light and electrons,\cite{LWuni,LWqed} we
assume that the space is formed by a collection of qubits and the qubits form a
string-net condensed state.  Light waves are collective motions of the
string-nets, and an electron corresponds to one end of string.  Such a
string-net unification of light and electrons has a  falsifiable prediction:
\emph{all fermionic excitations must carry some gauge
charges}.\cite{LWsta,LWstrnet}

The $U(1)\times SU(2) \times SU(3)$ standard model for elementary particles
contains fermionic excitations (such as neutrons and neutrinos) that do not
carry any $U(1)\times SU(2) \times SU(3)$ gauge charge.  So according to the
string-net theory, the $U(1)\times SU(2) \times SU(3)$ standard model is
incomplete.  According to the string-net theory, our universe not only have
$U(1)\times SU(2) \times SU(3)$ gauge theory, it must also contain other gauge
theories.  Those additional  gauge theories may have a gauge group of $Z_2$ or
other discrete groups.  Those extra discrete gauge theories will lead to new
cosmic strings which will appear in very early universe.

\section{A new chapter in physics}

Our world is rich and complex. When we discover the inner working of our world
and try to describe it, we ofter find that we need to invent new mathematical
language describe our understanding and insight.  For example, when Newton
discovered his law of mechanics, the proper mathematical language was not
invented yet.  Newton (and Leibniz) had to develop calculus in order to
formulate the law of mechanics.  For a long time, we tried to use theory of
mechanics and calculus to understand everything in our world.

As another example, when Einstein discovered the general equivalence principle
to describe gravity, he needed a  mathematical language to describe his theory.
In this case, the needed mathematics, Riemannian geometry, had been developed,
which leaded to the theory of general relativity.  Following the idea of
general relativity, we developed the gauge theory. Both general relativity and
gauge theory can be described by the mathematics of fiber bundles.  Those
advances led to a beautiful geometric understanding of our world based on
quantum field theory, and we tried to  understand everything in our world in
term of quantum field theory.

Now, I feel that we are at another turning point.  In a study of quantum
matter, we find that long-range entanglement can give rise to many new quantum
phases.  So long-range entanglement is a natural phenomenon that can happen in
our world. They greatly expand our understanding of possible quantum phases,
and bring the research of quantum matter to a whole new level.  To gain a
systematic understanding of new quantum phases and long-range entanglement, we
like to know what mathematical language should we use to describe long-range
entanglement?  The answer is not totally clear. But early studies suggest that
tensor category
\cite{KW9327,FNS0428,LWstrnet,RSW0777,W150605768,BBC1440,LW150704673,LW160205946}
and group cohomology \cite{CGL1314,CGL1204} should be a part of the
mathematical frame work that describes long-range entanglement.  The further
progresses in this direction will lead to a comprehensive understanding of
long-range entanglement and topological  quantum matter.

However, what is really exciting in the study of quantum matter is that it
might lead to a whole new point of view of our world. This is because
long-range entanglement can give rise to both gauge interactions and Fermi
statistics.  In contrast, the geometric point of view can only lead to gauge
interactions. So maybe we should not use geometric pictures, based on fields
and fiber bundles, to understand our world.  Maybe we should use entanglement
pictures to understand our world.  This way, we can get both gauge interactions
and fermions from a single origin -- qubits.  We may live in a truly quantum
world.  So, quantum entanglement represents a new chapter in physics.

\bibliography{../../bib/wencross,../../bib/all,../../bib/publst} 

\begin{thebibliography}{62}%
\makeatletter
\providecommand \@ifxundefined [1]{%
 \@ifx{#1\undefined}
}%
\providecommand \@ifnum [1]{%
 \ifnum #1\expandafter \@firstoftwo
 \else \expandafter \@secondoftwo
 \fi
}%
\providecommand \@ifx [1]{%
 \ifx #1\expandafter \@firstoftwo
 \else \expandafter \@secondoftwo
 \fi
}%
\providecommand \natexlab [1]{#1}%
\providecommand \enquote  [1]{``#1''}%
\providecommand \bibnamefont  [1]{#1}%
\providecommand \bibfnamefont [1]{#1}%
\providecommand \citenamefont [1]{#1}%
\providecommand \href@noop [0]{\@secondoftwo}%
\providecommand \href [0]{\begingroup \@sanitize@url \@href}%
\providecommand \@href[1]{\@@startlink{#1}\@@href}%
\providecommand \@@href[1]{\endgroup#1\@@endlink}%
\providecommand \@sanitize@url [0]{\catcode `\\12\catcode `\$12\catcode
  `\&12\catcode `\#12\catcode `\^12\catcode `\_12\catcode `\%12\relax}%
\providecommand \@@startlink[1]{}%
\providecommand \@@endlink[0]{}%
\providecommand \url  [0]{\begingroup\@sanitize@url \@url }%
\providecommand \@url [1]{\endgroup\@href {#1}{\urlprefix }}%
\providecommand \urlprefix  [0]{URL }%
\providecommand \Eprint [0]{\href }%
\providecommand \doibase [0]{http://dx.doi.org/}%
\providecommand \selectlanguage [0]{\@gobble}%
\providecommand \bibinfo  [0]{\@secondoftwo}%
\providecommand \bibfield  [0]{\@secondoftwo}%
\providecommand \translation [1]{[#1]}%
\providecommand \BibitemOpen [0]{}%
\providecommand \bibitemStop [0]{}%
\providecommand \bibitemNoStop [0]{.\EOS\space}%
\providecommand \EOS [0]{\spacefactor3000\relax}%
\providecommand \BibitemShut  [1]{\csname bibitem#1\endcsname}%
\let\auto@bib@innerbib\@empty
\bibitem [{\citenamefont {Foerster}\ \emph {et~al.}(1980)\citenamefont
  {Foerster}, \citenamefont {Nielsen},\ and\ \citenamefont
  {Ninomiya}}]{FNN8035}%
  \BibitemOpen
  \bibfield  {author} {\bibinfo {author} {\bibfnamefont {D.}~\bibnamefont
  {Foerster}}, \bibinfo {author} {\bibfnamefont {H.~B.}\ \bibnamefont
  {Nielsen}}, \ and\ \bibinfo {author} {\bibfnamefont {M.}~\bibnamefont
  {Ninomiya}},\ }\href@noop {} {\bibfield  {journal} {\bibinfo  {journal}
  {Phys. Lett. B}\ }\textbf {\bibinfo {volume} {94}},\ \bibinfo {pages} {135}
  (\bibinfo {year} {1980})}\BibitemShut {NoStop}%
\bibitem [{\citenamefont {Wen}(2003{\natexlab{a}})}]{Wqoem}%
  \BibitemOpen
  \bibfield  {author} {\bibinfo {author} {\bibfnamefont {X.-G.}\ \bibnamefont
  {Wen}},\ }\href@noop {} {\bibfield  {journal} {\bibinfo  {journal} {Phys.
  Rev. D}\ }\textbf {\bibinfo {volume} {68}},\ \bibinfo {pages} {065003}
  (\bibinfo {year} {2003}{\natexlab{a}})},\ \Eprint
  {http://arxiv.org/abs/hep-th/0302201} {hep-th/0302201} \BibitemShut {NoStop}%
\bibitem [{\citenamefont {Levin}\ and\ \citenamefont {Wen}(2006)}]{LWqed}%
  \BibitemOpen
  \bibfield  {author} {\bibinfo {author} {\bibfnamefont {M.}~\bibnamefont
  {Levin}}\ and\ \bibinfo {author} {\bibfnamefont {X.-G.}\ \bibnamefont
  {Wen}},\ }\href@noop {} {\bibfield  {journal} {\bibinfo  {journal} {Phys.
  Rev. B}\ }\textbf {\bibinfo {volume} {73}},\ \bibinfo {pages} {035122}
  (\bibinfo {year} {2006})},\ \Eprint {http://arxiv.org/abs/hep-th/0507118}
  {hep-th/0507118} \BibitemShut {NoStop}%
\bibitem [{\citenamefont {Gu}\ and\ \citenamefont {Wen}(2012)}]{GW1290}%
  \BibitemOpen
  \bibfield  {author} {\bibinfo {author} {\bibfnamefont {Z.-C.}\ \bibnamefont
  {Gu}}\ and\ \bibinfo {author} {\bibfnamefont {X.-G.}\ \bibnamefont {Wen}},\
  }\href@noop {} {\bibfield  {journal} {\bibinfo  {journal} {Nucl. Phys. B}\
  }\textbf {\bibinfo {volume} {863}},\ \bibinfo {pages} {90} (\bibinfo {year}
  {2012})},\ \Eprint {http://arxiv.org/abs/arXiv:0907.1203} {arXiv:0907.1203}
  \BibitemShut {NoStop}%
\bibitem [{\citenamefont {Wen}(2013{\natexlab{a}})}]{W1281}%
  \BibitemOpen
  \bibfield  {author} {\bibinfo {author} {\bibfnamefont {X.-G.}\ \bibnamefont
  {Wen}},\ }\href {\doibase 10.1155/2013/198710} {\bibfield  {journal}
  {\bibinfo  {journal} {ISRN Condensed Matter Physics}\ }\textbf {\bibinfo
  {volume} {2013}},\ \bibinfo {pages} {198710} (\bibinfo {year}
  {2013}{\natexlab{a}})},\ \Eprint {http://arxiv.org/abs/arXiv:1210.1281}
  {arXiv:1210.1281} \BibitemShut {NoStop}%
\bibitem [{\citenamefont {Wen}(2013{\natexlab{b}})}]{W1301}%
  \BibitemOpen
  \bibfield  {author} {\bibinfo {author} {\bibfnamefont {X.-G.}\ \bibnamefont
  {Wen}},\ }\href@noop {} {\bibfield  {journal} {\bibinfo  {journal} {Chin.
  Phys. Lett.}\ }\textbf {\bibinfo {volume} {30}},\ \bibinfo {pages} {111101}
  (\bibinfo {year} {2013}{\natexlab{b}})},\ \Eprint
  {http://arxiv.org/abs/arXiv:1305.1045} {arXiv:1305.1045} \BibitemShut
  {NoStop}%
\bibitem [{\citenamefont {{You}}\ and\ \citenamefont
  {{Xu}}(2015)}]{YX14124784}%
  \BibitemOpen
  \bibfield  {author} {\bibinfo {author} {\bibfnamefont {Y.-Z.}\ \bibnamefont
  {{You}}}\ and\ \bibinfo {author} {\bibfnamefont {C.}~\bibnamefont {{Xu}}},\
  }\href {\doibase 10.1103/PhysRevB.91.125147} {\bibfield  {journal} {\bibinfo
  {journal} {\prb}\ }\textbf {\bibinfo {volume} {91}},\ \bibinfo {pages}
  {125147} (\bibinfo {year} {2015})},\ \Eprint
  {http://arxiv.org/abs/arXiv:1412.4784} {arXiv:1412.4784} \BibitemShut
  {NoStop}%
\bibitem [{\citenamefont {{Zeng}}\ \emph {et~al.}(2015)\citenamefont {{Zeng}},
  \citenamefont {{Chen}}, \citenamefont {{Zhou}},\ and\ \citenamefont
  {{Wen}}}]{ZW150802595}%
  \BibitemOpen
  \bibfield  {author} {\bibinfo {author} {\bibfnamefont {B.}~\bibnamefont
  {{Zeng}}}, \bibinfo {author} {\bibfnamefont {X.}~\bibnamefont {{Chen}}},
  \bibinfo {author} {\bibfnamefont {D.-L.}\ \bibnamefont {{Zhou}}}, \ and\
  \bibinfo {author} {\bibfnamefont {X.-G.}\ \bibnamefont {{Wen}}},\ }\href@noop
  {} {\  (\bibinfo {year} {2015})},\ \Eprint
  {http://arxiv.org/abs/arXiv:1508.02595} {arXiv:1508.02595} \BibitemShut
  {NoStop}%
\bibitem [{\citenamefont {Einstein}(1916)}]{E1669}%
  \BibitemOpen
  \bibfield  {author} {\bibinfo {author} {\bibfnamefont {A.}~\bibnamefont
  {Einstein}},\ }\href@noop {} {\bibfield  {journal} {\bibinfo  {journal}
  {Annalen der Physik}\ }\textbf {\bibinfo {volume} {49}},\ \bibinfo {pages}
  {769} (\bibinfo {year} {1916})}\BibitemShut {NoStop}%
\bibitem [{\citenamefont {Nordstrom}\ and\ \citenamefont {die
  Moglichkeit}(1914)}]{NM1404}%
  \BibitemOpen
  \bibfield  {author} {\bibinfo {author} {\bibfnamefont {G.}~\bibnamefont
  {Nordstrom}}\ and\ \bibinfo {author} {\bibfnamefont {U.}~\bibnamefont {die
  Moglichkeit}},\ }\href@noop {} {\bibfield  {journal} {\bibinfo  {journal}
  {Physik. Zeitschr.}\ }\textbf {\bibinfo {volume} {15}},\ \bibinfo {pages}
  {504} (\bibinfo {year} {1914})}\BibitemShut {NoStop}%
\bibitem [{\citenamefont {Kaluza}(1921)}]{K2166}%
  \BibitemOpen
  \bibfield  {author} {\bibinfo {author} {\bibfnamefont {T.}~\bibnamefont
  {Kaluza}},\ }\href@noop {} {\bibfield  {journal} {\bibinfo  {journal}
  {Sitzungsber. Preuss. Akad. Wiss. Berlin. (Math. Phys.)}\ ,\ \bibinfo {pages}
  {966}} (\bibinfo {year} {1921})}\BibitemShut {NoStop}%
\bibitem [{\citenamefont {Klein}(1926)}]{K2695}%
  \BibitemOpen
  \bibfield  {author} {\bibinfo {author} {\bibfnamefont {O.}~\bibnamefont
  {Klein}},\ }\href@noop {} {\bibfield  {journal} {\bibinfo  {journal} {Z.
  Phys.}\ }\textbf {\bibinfo {volume} {37}},\ \bibinfo {pages} {895} (\bibinfo
  {year} {1926})}\BibitemShut {NoStop}%
\bibitem [{\citenamefont {Wen}(2014)}]{qubit2014}%
  \BibitemOpen
  \bibfield  {author} {\bibinfo {author} {\bibfnamefont {X.-G.}\ \bibnamefont
  {Wen}},\ }\href {http://blog.sciencenet.cn/blog-1116346-782883.html}
  {\bibfield  {journal} {\bibinfo  {journal} {{It from qubit, not bit,
  http://blog.sciencenet.cn/blog-1116346-782883.html}}\ } (\bibinfo {year}
  {2014})}\BibitemShut {NoStop}%
\bibitem [{\citenamefont {Weyl}(1952)}]{Wey52}%
  \BibitemOpen
  \bibfield  {author} {\bibinfo {author} {\bibfnamefont {H.}~\bibnamefont
  {Weyl}},\ }\href@noop {} {\emph {\bibinfo {title} {Space, Time, Matter}}}\
  (\bibinfo  {publisher} {Dover},\ \bibinfo {year} {1952})\BibitemShut
  {NoStop}%
\bibitem [{\citenamefont {Pauli}(1941)}]{P4103}%
  \BibitemOpen
  \bibfield  {author} {\bibinfo {author} {\bibfnamefont {W.}~\bibnamefont
  {Pauli}},\ }\href@noop {} {\bibfield  {journal} {\bibinfo  {journal} {Rev.
  Mod. Phys.}\ }\textbf {\bibinfo {volume} {13}},\ \bibinfo {pages} {203}
  (\bibinfo {year} {1941})}\BibitemShut {NoStop}%
\bibitem [{\citenamefont {Yang}\ and\ \citenamefont {Mills}(1954)}]{YM5491}%
  \BibitemOpen
  \bibfield  {author} {\bibinfo {author} {\bibfnamefont {C.~N.}\ \bibnamefont
  {Yang}}\ and\ \bibinfo {author} {\bibfnamefont {R.~L.}\ \bibnamefont
  {Mills}},\ }\href@noop {} {\bibfield  {journal} {\bibinfo  {journal} {Phys.
  Rev.}\ }\textbf {\bibinfo {volume} {96}},\ \bibinfo {pages} {191} (\bibinfo
  {year} {1954})}\BibitemShut {NoStop}%
\bibitem [{\citenamefont {Fermi}(1926)}]{F2602}%
  \BibitemOpen
  \bibfield  {author} {\bibinfo {author} {\bibfnamefont {E.}~\bibnamefont
  {Fermi}},\ }\href@noop {} {\bibfield  {journal} {\bibinfo  {journal} {Z.
  Phys.}\ }\textbf {\bibinfo {volume} {36}},\ \bibinfo {pages} {902} (\bibinfo
  {year} {1926})}\BibitemShut {NoStop}%
\bibitem [{\citenamefont {Dirac}(1926)}]{D2661}%
  \BibitemOpen
  \bibfield  {author} {\bibinfo {author} {\bibfnamefont {P.~A.~M.}\
  \bibnamefont {Dirac}},\ }\href@noop {} {\bibfield  {journal} {\bibinfo
  {journal} {Proc. Roy. Soc.}\ }\textbf {\bibinfo {volume} {A112}},\ \bibinfo
  {pages} {661} (\bibinfo {year} {1926})}\BibitemShut {NoStop}%
\bibitem [{\citenamefont {Gross}\ and\ \citenamefont {Wilczek}(1973)}]{GW7343}%
  \BibitemOpen
  \bibfield  {author} {\bibinfo {author} {\bibfnamefont {D.~J.}\ \bibnamefont
  {Gross}}\ and\ \bibinfo {author} {\bibfnamefont {F.}~\bibnamefont
  {Wilczek}},\ }\href@noop {} {\bibfield  {journal} {\bibinfo  {journal} {Phys.
  Rev. Lett.}\ }\textbf {\bibinfo {volume} {30}},\ \bibinfo {pages} {1343}
  (\bibinfo {year} {1973})}\BibitemShut {NoStop}%
\bibitem [{\citenamefont {Politzer}(1973)}]{P7346}%
  \BibitemOpen
  \bibfield  {author} {\bibinfo {author} {\bibfnamefont {H.~D.}\ \bibnamefont
  {Politzer}},\ }\href@noop {} {\bibfield  {journal} {\bibinfo  {journal}
  {Phys. Rev. Lett.}\ }\textbf {\bibinfo {volume} {30}},\ \bibinfo {pages}
  {1346} (\bibinfo {year} {1973})}\BibitemShut {NoStop}%
\bibitem [{\citenamefont {Lee}\ and\ \citenamefont {Yang}(1956)}]{LY5654}%
  \BibitemOpen
  \bibfield  {author} {\bibinfo {author} {\bibfnamefont {T.~D.}\ \bibnamefont
  {Lee}}\ and\ \bibinfo {author} {\bibfnamefont {C.~N.}\ \bibnamefont {Yang}},\
  }\href@noop {} {\bibfield  {journal} {\bibinfo  {journal} {Phys. Rev.}\
  }\textbf {\bibinfo {volume} {104}},\ \bibinfo {pages} {254} (\bibinfo {year}
  {1956})}\BibitemShut {NoStop}%
\bibitem [{\citenamefont {Wu}\ \emph {et~al.}(1957)\citenamefont {Wu} \emph
  {et~al.}}]{Wo5713}%
  \BibitemOpen
  \bibfield  {author} {\bibinfo {author} {\bibfnamefont {C.~S.}\ \bibnamefont
  {Wu}} \emph {et~al.},\ }\href@noop {} {\bibfield  {journal} {\bibinfo
  {journal} {Phys. Rev.}\ }\textbf {\bibinfo {volume} {105}},\ \bibinfo {pages}
  {1413} (\bibinfo {year} {1957})}\BibitemShut {NoStop}%
\bibitem [{\citenamefont {Einstein}(1905)}]{E0591}%
  \BibitemOpen
  \bibfield  {author} {\bibinfo {author} {\bibfnamefont {A.}~\bibnamefont
  {Einstein}},\ }\href@noop {} {\bibfield  {journal} {\bibinfo  {journal}
  {Annalen der Physik}\ }\textbf {\bibinfo {volume} {17}},\ \bibinfo {pages}
  {891} (\bibinfo {year} {1905})}\BibitemShut {NoStop}%
\bibitem [{\citenamefont {Tsui}\ \emph {et~al.}(1982)\citenamefont {Tsui},
  \citenamefont {Stormer},\ and\ \citenamefont {Gossard}}]{TSG8259}%
  \BibitemOpen
  \bibfield  {author} {\bibinfo {author} {\bibfnamefont {D.~C.}\ \bibnamefont
  {Tsui}}, \bibinfo {author} {\bibfnamefont {H.~L.}\ \bibnamefont {Stormer}}, \
  and\ \bibinfo {author} {\bibfnamefont {A.~C.}\ \bibnamefont {Gossard}},\
  }\href@noop {} {\bibfield  {journal} {\bibinfo  {journal} {Phys. Rev. Lett.}\
  }\textbf {\bibinfo {volume} {48}},\ \bibinfo {pages} {1559} (\bibinfo {year}
  {1982})}\BibitemShut {NoStop}%
\bibitem [{\citenamefont {Chen}\ \emph {et~al.}(2010)\citenamefont {Chen},
  \citenamefont {Gu},\ and\ \citenamefont {Wen}}]{CGW1038}%
  \BibitemOpen
  \bibfield  {author} {\bibinfo {author} {\bibfnamefont {X.}~\bibnamefont
  {Chen}}, \bibinfo {author} {\bibfnamefont {Z.-C.}\ \bibnamefont {Gu}}, \ and\
  \bibinfo {author} {\bibfnamefont {X.-G.}\ \bibnamefont {Wen}},\ }\href@noop
  {} {\bibfield  {journal} {\bibinfo  {journal} {Phys. Rev. B}\ }\textbf
  {\bibinfo {volume} {82}},\ \bibinfo {pages} {155138} (\bibinfo {year}
  {2010})},\ \Eprint {http://arxiv.org/abs/arXiv:1004.3835} {arXiv:1004.3835}
  \BibitemShut {NoStop}%
\bibitem [{\citenamefont {Wen}(1989)}]{Wtop}%
  \BibitemOpen
  \bibfield  {author} {\bibinfo {author} {\bibfnamefont {X.-G.}\ \bibnamefont
  {Wen}},\ }\href@noop {} {\bibfield  {journal} {\bibinfo  {journal} {Phys.
  Rev. B}\ }\textbf {\bibinfo {volume} {40}},\ \bibinfo {pages} {7387}
  (\bibinfo {year} {1989})}\BibitemShut {NoStop}%
\bibitem [{\citenamefont {Wen}(1990)}]{Wrig}%
  \BibitemOpen
  \bibfield  {author} {\bibinfo {author} {\bibfnamefont {X.-G.}\ \bibnamefont
  {Wen}},\ }\href@noop {} {\bibfield  {journal} {\bibinfo  {journal} {Int. J.
  Mod. Phys. B}\ }\textbf {\bibinfo {volume} {4}},\ \bibinfo {pages} {239}
  (\bibinfo {year} {1990})}\BibitemShut {NoStop}%
\bibitem [{\citenamefont {Wen}(2002)}]{Wlight}%
  \BibitemOpen
  \bibfield  {author} {\bibinfo {author} {\bibfnamefont {X.-G.}\ \bibnamefont
  {Wen}},\ }\href@noop {} {\bibfield  {journal} {\bibinfo  {journal} {Phys.
  Rev. Lett.}\ }\textbf {\bibinfo {volume} {88}},\ \bibinfo {pages} {11602}
  (\bibinfo {year} {2002})},\ \Eprint {http://arxiv.org/abs/hep-th/01090120}
  {hep-th/01090120} \BibitemShut {NoStop}%
\bibitem [{\citenamefont {Gu}\ and\ \citenamefont {Wen}(2006)}]{GW0600}%
  \BibitemOpen
  \bibfield  {author} {\bibinfo {author} {\bibfnamefont {Z.-C.}\ \bibnamefont
  {Gu}}\ and\ \bibinfo {author} {\bibfnamefont {X.-G.}\ \bibnamefont {Wen}},\
  }\href@noop {} {\  (\bibinfo {year} {2006})},\ \Eprint
  {http://arxiv.org/abs/gr-qc/0606100} {gr-qc/0606100} \BibitemShut {NoStop}%
\bibitem [{\citenamefont {Levin}\ and\ \citenamefont
  {Wen}(2005{\natexlab{a}})}]{LWstrnet}%
  \BibitemOpen
  \bibfield  {author} {\bibinfo {author} {\bibfnamefont {M.}~\bibnamefont
  {Levin}}\ and\ \bibinfo {author} {\bibfnamefont {X.-G.}\ \bibnamefont
  {Wen}},\ }\href@noop {} {\bibfield  {journal} {\bibinfo  {journal} {Phys.
  Rev. B}\ }\textbf {\bibinfo {volume} {71}},\ \bibinfo {pages} {045110}
  (\bibinfo {year} {2005}{\natexlab{a}})},\ \Eprint
  {http://arxiv.org/abs/cond-mat/0404617} {cond-mat/0404617} \BibitemShut
  {NoStop}%
\bibitem [{\citenamefont {Levin}\ and\ \citenamefont
  {Wen}(2005{\natexlab{b}})}]{LWuni}%
  \BibitemOpen
  \bibfield  {author} {\bibinfo {author} {\bibfnamefont {M.~A.}\ \bibnamefont
  {Levin}}\ and\ \bibinfo {author} {\bibfnamefont {X.-G.}\ \bibnamefont
  {Wen}},\ }\href@noop {} {\bibfield  {journal} {\bibinfo  {journal} {Rev. Mod.
  Phys.}\ }\textbf {\bibinfo {volume} {77}},\ \bibinfo {pages} {871} (\bibinfo
  {year} {2005}{\natexlab{b}})},\ \Eprint
  {http://arxiv.org/abs/cond-mat/0407140} {cond-mat/0407140} \BibitemShut
  {NoStop}%
\bibitem [{\citenamefont {Wen}(2003{\natexlab{b}})}]{Walight}%
  \BibitemOpen
  \bibfield  {author} {\bibinfo {author} {\bibfnamefont {X.-G.}\ \bibnamefont
  {Wen}},\ }\href@noop {} {\bibfield  {journal} {\bibinfo  {journal} {Phys.
  Rev. B}\ }\textbf {\bibinfo {volume} {68}},\ \bibinfo {pages} {115413}
  (\bibinfo {year} {2003}{\natexlab{b}})},\ \Eprint
  {http://arxiv.org/abs/cond-mat/0210040} {cond-mat/0210040} \BibitemShut
  {NoStop}%
\bibitem [{\citenamefont {Moessner}\ and\ \citenamefont
  {Sondhi}(2003)}]{MS0312}%
  \BibitemOpen
  \bibfield  {author} {\bibinfo {author} {\bibfnamefont {R.}~\bibnamefont
  {Moessner}}\ and\ \bibinfo {author} {\bibfnamefont {S.~L.}\ \bibnamefont
  {Sondhi}},\ }\href@noop {} {\bibfield  {journal} {\bibinfo  {journal} {Phys.
  Rev. B}\ }\textbf {\bibinfo {volume} {68}},\ \bibinfo {pages} {184512}
  (\bibinfo {year} {2003})}\BibitemShut {NoStop}%
\bibitem [{\citenamefont {Hermele}\ \emph {et~al.}(2004)\citenamefont
  {Hermele}, \citenamefont {Fisher},\ and\ \citenamefont {Balents}}]{HFB0404}%
  \BibitemOpen
  \bibfield  {author} {\bibinfo {author} {\bibfnamefont {M.}~\bibnamefont
  {Hermele}}, \bibinfo {author} {\bibfnamefont {M.~P.~A.}\ \bibnamefont
  {Fisher}}, \ and\ \bibinfo {author} {\bibfnamefont {L.}~\bibnamefont
  {Balents}},\ }\href@noop {} {\bibfield  {journal} {\bibinfo  {journal} {Phys.
  Rev. B}\ }\textbf {\bibinfo {volume} {69}},\ \bibinfo {pages} {064404}
  (\bibinfo {year} {2004})}\BibitemShut {NoStop}%
\bibitem [{\citenamefont {Levin}\ and\ \citenamefont {Wen}(2003)}]{LWsta}%
  \BibitemOpen
  \bibfield  {author} {\bibinfo {author} {\bibfnamefont {M.}~\bibnamefont
  {Levin}}\ and\ \bibinfo {author} {\bibfnamefont {X.-G.}\ \bibnamefont
  {Wen}},\ }\href@noop {} {\bibfield  {journal} {\bibinfo  {journal} {Phys.
  Rev. B}\ }\textbf {\bibinfo {volume} {67}},\ \bibinfo {pages} {245316}
  (\bibinfo {year} {2003})},\ \Eprint {http://arxiv.org/abs/cond-mat/0302460}
  {cond-mat/0302460} \BibitemShut {NoStop}%
\bibitem [{\citenamefont {D'Adda}\ \emph {et~al.}(1978)\citenamefont {D'Adda},
  \citenamefont {Vecchia},\ and\ \citenamefont {L{\"u}scher}}]{DDL7863}%
  \BibitemOpen
  \bibfield  {author} {\bibinfo {author} {\bibfnamefont {A.}~\bibnamefont
  {D'Adda}}, \bibinfo {author} {\bibfnamefont {P.~D.}\ \bibnamefont {Vecchia}},
  \ and\ \bibinfo {author} {\bibfnamefont {M.}~\bibnamefont {L{\"u}scher}},\
  }\href@noop {} {\bibfield  {journal} {\bibinfo  {journal} {Nucl. Phys. B}\
  }\textbf {\bibinfo {volume} {146}},\ \bibinfo {pages} {63} (\bibinfo {year}
  {1978})}\BibitemShut {NoStop}%
\bibitem [{\citenamefont {Witten}(1979)}]{W7985}%
  \BibitemOpen
  \bibfield  {author} {\bibinfo {author} {\bibfnamefont {E.}~\bibnamefont
  {Witten}},\ }\href@noop {} {\bibfield  {journal} {\bibinfo  {journal} {Nucl.
  Phys. B}\ }\textbf {\bibinfo {volume} {149}},\ \bibinfo {pages} {285}
  (\bibinfo {year} {1979})}\BibitemShut {NoStop}%
\bibitem [{\citenamefont {Baskaran}\ and\ \citenamefont
  {Anderson}(1988)}]{BA8880}%
  \BibitemOpen
  \bibfield  {author} {\bibinfo {author} {\bibfnamefont {G.}~\bibnamefont
  {Baskaran}}\ and\ \bibinfo {author} {\bibfnamefont {P.~W.}\ \bibnamefont
  {Anderson}},\ }\href@noop {} {\bibfield  {journal} {\bibinfo  {journal}
  {Phys. Rev. B}\ }\textbf {\bibinfo {volume} {37}},\ \bibinfo {pages} {580}
  (\bibinfo {year} {1988})}\BibitemShut {NoStop}%
\bibitem [{\citenamefont {Affleck}\ and\ \citenamefont
  {Marston}(1988)}]{AM8874}%
  \BibitemOpen
  \bibfield  {author} {\bibinfo {author} {\bibfnamefont {I.}~\bibnamefont
  {Affleck}}\ and\ \bibinfo {author} {\bibfnamefont {J.~B.}\ \bibnamefont
  {Marston}},\ }\href@noop {} {\bibfield  {journal} {\bibinfo  {journal} {Phys.
  Rev. B}\ }\textbf {\bibinfo {volume} {37}},\ \bibinfo {pages} {3774}
  (\bibinfo {year} {1988})}\BibitemShut {NoStop}%
\bibitem [{\citenamefont {Wilson}(1974)}]{W7445}%
  \BibitemOpen
  \bibfield  {author} {\bibinfo {author} {\bibfnamefont {K.~G.}\ \bibnamefont
  {Wilson}},\ }\href@noop {} {\bibfield  {journal} {\bibinfo  {journal} {Phys.
  Rev. D}\ }\textbf {\bibinfo {volume} {10}},\ \bibinfo {pages} {2445}
  (\bibinfo {year} {1974})}\BibitemShut {NoStop}%
\bibitem [{\citenamefont {Kogut}\ and\ \citenamefont
  {Susskind}(1975)}]{KS7595}%
  \BibitemOpen
  \bibfield  {author} {\bibinfo {author} {\bibfnamefont {J.}~\bibnamefont
  {Kogut}}\ and\ \bibinfo {author} {\bibfnamefont {L.}~\bibnamefont
  {Susskind}},\ }\href@noop {} {\bibfield  {journal} {\bibinfo  {journal}
  {Phys. Rev. D}\ }\textbf {\bibinfo {volume} {11}},\ \bibinfo {pages} {395}
  (\bibinfo {year} {1975})}\BibitemShut {NoStop}%
\bibitem [{\citenamefont {Banks}\ \emph {et~al.}(1977)\citenamefont {Banks},
  \citenamefont {Myerson},\ and\ \citenamefont {Kogut}}]{BMK7793}%
  \BibitemOpen
  \bibfield  {author} {\bibinfo {author} {\bibfnamefont {T.}~\bibnamefont
  {Banks}}, \bibinfo {author} {\bibfnamefont {R.}~\bibnamefont {Myerson}}, \
  and\ \bibinfo {author} {\bibfnamefont {J.~B.}\ \bibnamefont {Kogut}},\
  }\href@noop {} {\bibfield  {journal} {\bibinfo  {journal} {Nucl. Phys. B}\
  }\textbf {\bibinfo {volume} {129}},\ \bibinfo {pages} {493} (\bibinfo {year}
  {1977})}\BibitemShut {NoStop}%
\bibitem [{\citenamefont {Kogut}(1979)}]{K7959}%
  \BibitemOpen
  \bibfield  {author} {\bibinfo {author} {\bibfnamefont {J.~B.}\ \bibnamefont
  {Kogut}},\ }\href {\doibase 10.1103/RevModPhys.51.659} {\bibfield  {journal}
  {\bibinfo  {journal} {Rev. Mod. Phys.}\ }\textbf {\bibinfo {volume} {51}},\
  \bibinfo {pages} {659 } (\bibinfo {year} {1979})}\BibitemShut {NoStop}%
\bibitem [{\citenamefont {Savit}(1980)}]{S8053}%
  \BibitemOpen
  \bibfield  {author} {\bibinfo {author} {\bibfnamefont {R.}~\bibnamefont
  {Savit}},\ }\href@noop {} {\bibfield  {journal} {\bibinfo  {journal} {Rev.
  Mod. Phys.}\ }\textbf {\bibinfo {volume} {52}},\ \bibinfo {pages} {453}
  (\bibinfo {year} {1980})}\BibitemShut {NoStop}%
\bibitem [{\citenamefont {Hastings}\ and\ \citenamefont {Wen}(2005)}]{HWcnt}%
  \BibitemOpen
  \bibfield  {author} {\bibinfo {author} {\bibfnamefont {M.~B.}\ \bibnamefont
  {Hastings}}\ and\ \bibinfo {author} {\bibfnamefont {X.-G.}\ \bibnamefont
  {Wen}},\ }\href@noop {} {\bibfield  {journal} {\bibinfo  {journal} {Phys.
  Rev. B}\ }\textbf {\bibinfo {volume} {72}},\ \bibinfo {pages} {045141}
  (\bibinfo {year} {2005})},\ \Eprint {http://arxiv.org/abs/cond-mat/0503554}
  {cond-mat/0503554} \BibitemShut {NoStop}%
\bibitem [{\citenamefont {Motrunich}\ and\ \citenamefont
  {Senthil}(2002)}]{MS0204}%
  \BibitemOpen
  \bibfield  {author} {\bibinfo {author} {\bibfnamefont {O.~I.}\ \bibnamefont
  {Motrunich}}\ and\ \bibinfo {author} {\bibfnamefont {T.}~\bibnamefont
  {Senthil}},\ }\href@noop {} {\bibfield  {journal} {\bibinfo  {journal} {Phys.
  Rev. Lett.}\ }\textbf {\bibinfo {volume} {89}},\ \bibinfo {pages} {277004}
  (\bibinfo {year} {2002})}\BibitemShut {NoStop}%
\bibitem [{\citenamefont {Leinaas}\ and\ \citenamefont
  {Myrheim}(1977)}]{LM7701}%
  \BibitemOpen
  \bibfield  {author} {\bibinfo {author} {\bibfnamefont {J.~M.}\ \bibnamefont
  {Leinaas}}\ and\ \bibinfo {author} {\bibfnamefont {J.}~\bibnamefont
  {Myrheim}},\ }\href@noop {} {\bibfield  {journal} {\bibinfo  {journal} {Il
  Nuovo Cimento}\ }\textbf {\bibinfo {volume} {37B}},\ \bibinfo {pages} {1}
  (\bibinfo {year} {1977})}\BibitemShut {NoStop}%
\bibitem [{\citenamefont {Wilczek}(1982{\natexlab{a}})}]{W8257}%
  \BibitemOpen
  \bibfield  {author} {\bibinfo {author} {\bibfnamefont {F.}~\bibnamefont
  {Wilczek}},\ }\href@noop {} {\bibfield  {journal} {\bibinfo  {journal} {Phys.
  Rev. Lett.}\ }\textbf {\bibinfo {volume} {49}},\ \bibinfo {pages} {957}
  (\bibinfo {year} {1982}{\natexlab{a}})}\BibitemShut {NoStop}%
\bibitem [{\citenamefont {Tamm}(1931)}]{T3141}%
  \BibitemOpen
  \bibfield  {author} {\bibinfo {author} {\bibfnamefont {I.}~\bibnamefont
  {Tamm}},\ }\href@noop {} {\bibfield  {journal} {\bibinfo  {journal} {Z.
  Phys.}\ }\textbf {\bibinfo {volume} {71}},\ \bibinfo {pages} {141} (\bibinfo
  {year} {1931})}\BibitemShut {NoStop}%
\bibitem [{\citenamefont {Jackiw}\ and\ \citenamefont {Rebbi}(1976)}]{JR7616}%
  \BibitemOpen
  \bibfield  {author} {\bibinfo {author} {\bibfnamefont {R.}~\bibnamefont
  {Jackiw}}\ and\ \bibinfo {author} {\bibfnamefont {C.}~\bibnamefont {Rebbi}},\
  }\href {\doibase 10.1103/PhysRevLett.36.1116} {\bibfield  {journal} {\bibinfo
   {journal} {Phys. Rev. Lett.}\ }\textbf {\bibinfo {volume} {36}},\ \bibinfo
  {pages} {1116 } (\bibinfo {year} {1976})}\BibitemShut {NoStop}%
\bibitem [{\citenamefont {Wilczek}(1982{\natexlab{b}})}]{W8246}%
  \BibitemOpen
  \bibfield  {author} {\bibinfo {author} {\bibfnamefont {F.}~\bibnamefont
  {Wilczek}},\ }\href {\doibase 10.1103/PhysRevLett.48.1146} {\bibfield
  {journal} {\bibinfo  {journal} {Phys. Rev. Lett.}\ }\textbf {\bibinfo
  {volume} {48}},\ \bibinfo {pages} {1146 } (\bibinfo {year}
  {1982}{\natexlab{b}})}\BibitemShut {NoStop}%
\bibitem [{\citenamefont {Goldhaber}(1982)}]{G8205}%
  \BibitemOpen
  \bibfield  {author} {\bibinfo {author} {\bibfnamefont {A.~S.}\ \bibnamefont
  {Goldhaber}},\ }\href@noop {} {\bibfield  {journal} {\bibinfo  {journal}
  {Phys. Rev. Lett.}\ }\textbf {\bibinfo {volume} {49}},\ \bibinfo {pages} {905
  } (\bibinfo {year} {1982})}\BibitemShut {NoStop}%
\bibitem [{\citenamefont {Lechner}\ and\ \citenamefont
  {Marchetti}(2000)}]{LM0012}%
  \BibitemOpen
  \bibfield  {author} {\bibinfo {author} {\bibfnamefont {K.}~\bibnamefont
  {Lechner}}\ and\ \bibinfo {author} {\bibfnamefont {P.~A.}\ \bibnamefont
  {Marchetti}},\ }\href {\doibase 10.1088/1126-6708/2000/12/028} {\bibfield
  {journal} {\bibinfo  {journal} {Journal of High Energy Physics}\ }\textbf
  {\bibinfo {volume} {2000}},\ \bibinfo {pages} {12} (\bibinfo {year}
  {2000})},\ \Eprint {http://arxiv.org/abs/hep-th/0010291} {hep-th/0010291}
  \BibitemShut {NoStop}%
\bibitem [{\citenamefont {Keski-Vakkuri}\ and\ \citenamefont
  {Wen}(1993)}]{KW9327}%
  \BibitemOpen
  \bibfield  {author} {\bibinfo {author} {\bibfnamefont {E.}~\bibnamefont
  {Keski-Vakkuri}}\ and\ \bibinfo {author} {\bibfnamefont {X.-G.}\ \bibnamefont
  {Wen}},\ }\href@noop {} {\bibfield  {journal} {\bibinfo  {journal} {Int. J.
  Mod. Phys. B}\ }\textbf {\bibinfo {volume} {7}},\ \bibinfo {pages} {4227}
  (\bibinfo {year} {1993})}\BibitemShut {NoStop}%
\bibitem [{\citenamefont {Freedman}\ \emph {et~al.}(2004)\citenamefont
  {Freedman}, \citenamefont {Nayak}, \citenamefont {Shtengel}, \citenamefont
  {Walker},\ and\ \citenamefont {Wang}}]{FNS0428}%
  \BibitemOpen
  \bibfield  {author} {\bibinfo {author} {\bibfnamefont {M.}~\bibnamefont
  {Freedman}}, \bibinfo {author} {\bibfnamefont {C.}~\bibnamefont {Nayak}},
  \bibinfo {author} {\bibfnamefont {K.}~\bibnamefont {Shtengel}}, \bibinfo
  {author} {\bibfnamefont {K.}~\bibnamefont {Walker}}, \ and\ \bibinfo {author}
  {\bibfnamefont {Z.}~\bibnamefont {Wang}},\ }\href@noop {} {\bibfield
  {journal} {\bibinfo  {journal} {Ann. Phys. (NY)}\ }\textbf {\bibinfo {volume}
  {310}},\ \bibinfo {pages} {428} (\bibinfo {year} {2004})},\ \Eprint
  {http://arxiv.org/abs/cond-mat/0307511} {cond-mat/0307511} \BibitemShut
  {NoStop}%
\bibitem [{\citenamefont {{Rowell}}\ \emph {et~al.}(2009)\citenamefont
  {{Rowell}}, \citenamefont {{Stong}},\ and\ \citenamefont {{Wang}}}]{RSW0777}%
  \BibitemOpen
  \bibfield  {author} {\bibinfo {author} {\bibfnamefont {E.}~\bibnamefont
  {{Rowell}}}, \bibinfo {author} {\bibfnamefont {R.}~\bibnamefont {{Stong}}}, \
  and\ \bibinfo {author} {\bibfnamefont {Z.}~\bibnamefont {{Wang}}},\
  }\href@noop {} {\bibfield  {journal} {\bibinfo  {journal} {Comm. Math.
  Phys.}\ }\textbf {\bibinfo {volume} {292}},\ \bibinfo {pages} {343} (\bibinfo
  {year} {2009})},\ \Eprint {http://arxiv.org/abs/arXiv:0712.1377}
  {arXiv:0712.1377} \BibitemShut {NoStop}%
\bibitem [{\citenamefont {{Wen}}(2016)}]{W150605768}%
  \BibitemOpen
  \bibfield  {author} {\bibinfo {author} {\bibfnamefont {X.-G.}\ \bibnamefont
  {{Wen}}},\ }\href {\doibase 10.1093/nsr/nwv077} {\bibfield  {journal}
  {\bibinfo  {journal} {Natl. Sci. Rev.}\ }\textbf {\bibinfo {volume} {3}},\
  \bibinfo {pages} {68} (\bibinfo {year} {2016})},\ \Eprint
  {http://arxiv.org/abs/arXiv:1506.05768} {arXiv:1506.05768} \BibitemShut
  {NoStop}%
\bibitem [{\citenamefont {{Barkeshli}}\ \emph {et~al.}(2014)\citenamefont
  {{Barkeshli}}, \citenamefont {{Bonderson}}, \citenamefont {{Cheng}},\ and\
  \citenamefont {{Wang}}}]{BBC1440}%
  \BibitemOpen
  \bibfield  {author} {\bibinfo {author} {\bibfnamefont {M.}~\bibnamefont
  {{Barkeshli}}}, \bibinfo {author} {\bibfnamefont {P.}~\bibnamefont
  {{Bonderson}}}, \bibinfo {author} {\bibfnamefont {M.}~\bibnamefont
  {{Cheng}}}, \ and\ \bibinfo {author} {\bibfnamefont {Z.}~\bibnamefont
  {{Wang}}},\ }\href@noop {} {\  (\bibinfo {year} {2014})},\ \Eprint
  {http://arxiv.org/abs/arXiv:1410.4540} {arXiv:1410.4540} \BibitemShut
  {NoStop}%
\bibitem [{\citenamefont {{Lan}}\ \emph {et~al.}(2016)\citenamefont {{Lan}},
  \citenamefont {{Kong}},\ and\ \citenamefont {{Wen}}}]{LW150704673}%
  \BibitemOpen
  \bibfield  {author} {\bibinfo {author} {\bibfnamefont {T.}~\bibnamefont
  {{Lan}}}, \bibinfo {author} {\bibfnamefont {L.}~\bibnamefont {{Kong}}}, \
  and\ \bibinfo {author} {\bibfnamefont {X.-G.}\ \bibnamefont {{Wen}}},\ }\href
  {\doibase 10.1103/PhysRevB.94.155113} {\bibfield  {journal} {\bibinfo
  {journal} {\prb}\ }\textbf {\bibinfo {volume} {94}},\ \bibinfo {pages}
  {155113} (\bibinfo {year} {2016})},\ \Eprint
  {http://arxiv.org/abs/arXiv:1507.04673} {arXiv:1507.04673} \BibitemShut
  {NoStop}%
\bibitem [{\citenamefont {Lan}\ \emph {et~al.}(2016)\citenamefont {Lan},
  \citenamefont {Kong},\ and\ \citenamefont {Wen}}]{LW160205946}%
  \BibitemOpen
  \bibfield  {author} {\bibinfo {author} {\bibfnamefont {T.}~\bibnamefont
  {Lan}}, \bibinfo {author} {\bibfnamefont {L.}~\bibnamefont {Kong}}, \ and\
  \bibinfo {author} {\bibfnamefont {X.-G.}\ \bibnamefont {Wen}},\ }\href
  {\doibase 10.1103/PhysRevB.95.235140} {\bibfield  {journal} {\bibinfo
  {journal} {Phys. Rev. B}\ }\textbf {\bibinfo {volume} {95}},\ \bibinfo
  {pages} {235140} (\bibinfo {year} {2016})},\ \Eprint
  {http://arxiv.org/abs/arXiv:1602.05946} {arXiv:1602.05946} \BibitemShut
  {NoStop}%
\bibitem [{\citenamefont {Chen}\ \emph {et~al.}(2013)\citenamefont {Chen},
  \citenamefont {Gu}, \citenamefont {Liu},\ and\ \citenamefont
  {Wen}}]{CGL1314}%
  \BibitemOpen
  \bibfield  {author} {\bibinfo {author} {\bibfnamefont {X.}~\bibnamefont
  {Chen}}, \bibinfo {author} {\bibfnamefont {Z.-C.}\ \bibnamefont {Gu}},
  \bibinfo {author} {\bibfnamefont {Z.-X.}\ \bibnamefont {Liu}}, \ and\
  \bibinfo {author} {\bibfnamefont {X.-G.}\ \bibnamefont {Wen}},\ }\href@noop
  {} {\bibfield  {journal} {\bibinfo  {journal} {Phys. Rev. B}\ }\textbf
  {\bibinfo {volume} {87}},\ \bibinfo {pages} {155114} (\bibinfo {year}
  {2013})},\ \Eprint {http://arxiv.org/abs/arXiv:1106.4772} {arXiv:1106.4772}
  \BibitemShut {NoStop}%
\bibitem [{\citenamefont {Chen}\ \emph {et~al.}(2012)\citenamefont {Chen},
  \citenamefont {Gu}, \citenamefont {Liu},\ and\ \citenamefont
  {Wen}}]{CGL1204}%
  \BibitemOpen
  \bibfield  {author} {\bibinfo {author} {\bibfnamefont {X.}~\bibnamefont
  {Chen}}, \bibinfo {author} {\bibfnamefont {Z.-C.}\ \bibnamefont {Gu}},
  \bibinfo {author} {\bibfnamefont {Z.-X.}\ \bibnamefont {Liu}}, \ and\
  \bibinfo {author} {\bibfnamefont {X.-G.}\ \bibnamefont {Wen}},\ }\href@noop
  {} {\bibfield  {journal} {\bibinfo  {journal} {Science}\ }\textbf {\bibinfo
  {volume} {338}},\ \bibinfo {pages} {1604} (\bibinfo {year} {2012})},\ \Eprint
  {http://arxiv.org/abs/arXiv:1301.0861} {arXiv:1301.0861} \BibitemShut
  {NoStop}%
\end{thebibliography}%
\end{document}